\documentclass[usenatbib]{mnras}
\usepackage{mathptmx}
\usepackage{color}
\usepackage{graphics,graphicx,amssymb,amsmath,natbib}
\usepackage{mymacros}
\usepackage{epsfig}
\usepackage[usenames]{xcolor}
\usepackage{float}

\def\xs{{IRS~13E}}
\def\h30a{{H30$\alpha$}}
\def\pa{{Pa-$\alpha$}}
\def\kms{{\rm~km~s^{-1}}}
\def\alma{{\it ALMA}}

\begin{document}

\title{Colliding Winds in and around the Stellar Group IRS~13E at the Galactic Center}

\author[]{Q. Daniel Wang$^{1,2}$\thanks{E-mail:wqd@umass.edu}, Jun Li$^{1,2,3}$,
	Christopher M. P. Russell$^{2}$,  
	Jorge Cuadra$^{2}$\\ 
$^{1}$Department of Astronomy, University of Massachusetts,  Amherst, MA 01003, USA\\
$^{2}$Instituto de Astrof\'i­sica, Facultad de F\'i­sica, Pontificia Universidad Cat\'olica de Chile, 782-0436 Santiago, Chile\\
$^{3}$ Department of Astronomy, Beijing Normal University, Beijing 100875, China\\
}
\maketitle

\begin{abstract}
IRS~13E is an enigmatic compact group of massive stars located in projection only 3\farcs6 away from Sgr A*. This group has been suggested to be bounded by an intermediate-mass black hole (IMBH). We present a multi-wavelength study of the group and its interplay with the environment. Based on \chandra\ observations, we find the X-ray spectrum of IRS~13E can be well characterized by an optically thin thermal plasma. The emission peaks between two strongly mass-losing Wolf-Rayet stars of the group. These properties can be reasonably well reproduced by simulated colliding winds of these two stars. However, this scenario under-predicts the X-ray intensity in outer regions. The residual emission likely results from the ram-pressure confinement of the IRS~13E group wind by the ambient medium and is apparently associated with a shell-like warm gas structure seen in Pa-alpha and in {\sl ALMA} observations. These latter observations also show strongly peaked thermal emission with unusually large velocity spread between the two stars. These results indicate that the group is colliding with the bar of the dense cool gas mini-spiral around Sgr A*. The extended X-ray morphology of IRS~13E and its association with the bar further suggest that the group is physically much farther away than the projected distance from Sgr A*. The presence of an IMBH, while favorable to keep the stars bound together, is not necessary to explain the observed stellar and gas properties of IRS~13E.

\end{abstract}


\section{Introduction}

The Galactic center (GC) of the Milky Way is a unique site for the detailed study of a multitude of complex astrophysical phenomena that appear common to the nuclear regions of many galaxies.
Multiple stellar populations and structures have been identified around Sgr A* -- the supermassive black hole (SMBH) of our Galaxy~\citep[e.g.,][]{paumard06,bartko09,schodel14}. Massive stars, in particular, appear to form such perplexing structures as the clock-wise disk (CWD) and possibly a counter-clock-wise disk (CCWD). In addition, a compact group of massive stars, \xs, has long been recognized and may or may not belong to the CCWD~\citep{genzel10}. The formation and dynamics of these distinct massive star structures, as well as their interplay with the gaseous environment and Sgr A*, have been an active research topic for many years~\citep[e.g.,][]{nayakshin07,cuadra08}.

\begin{figure*} 
\centering
\includegraphics[width=0.95\linewidth,angle=0]{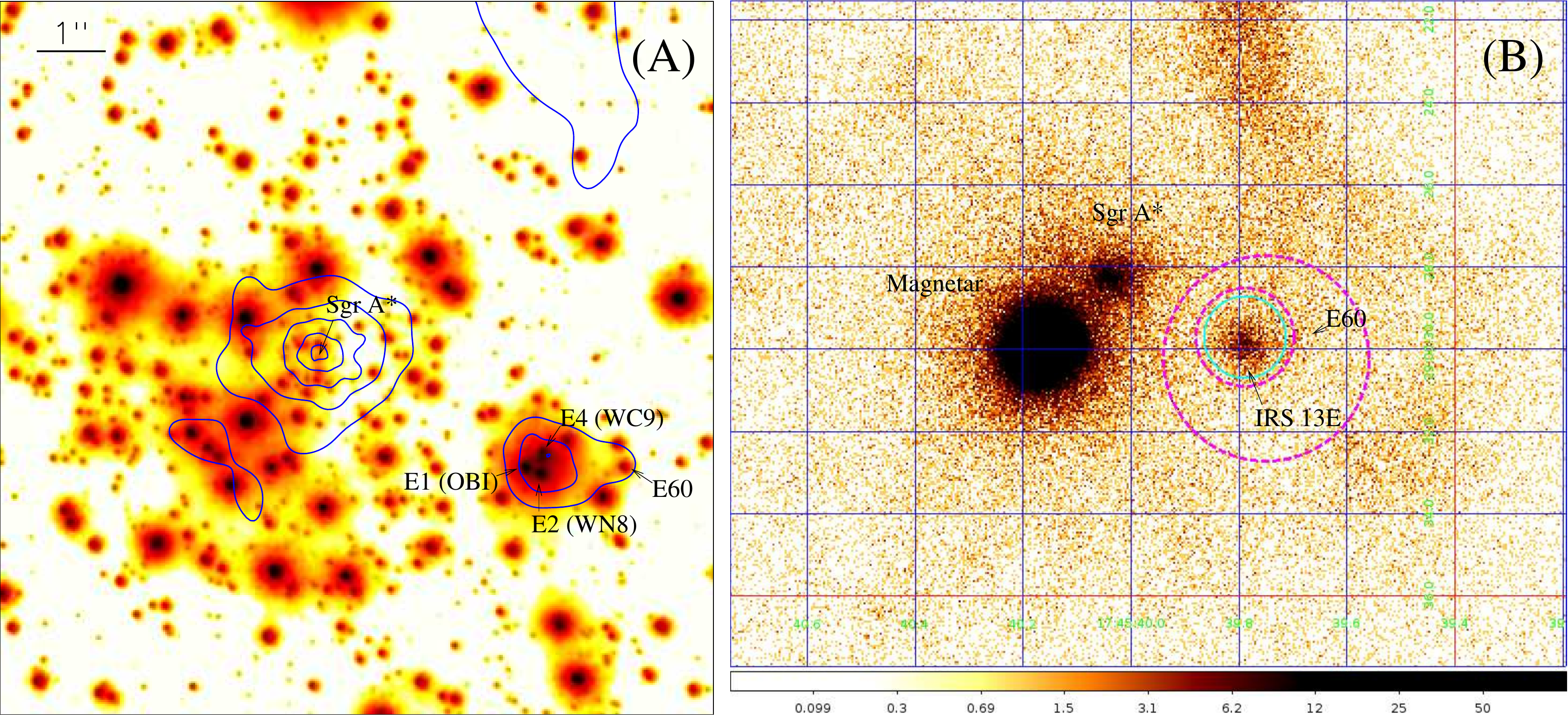}
\caption{
X-ray views of the \xs\ and Sgr A* region: (A) Overlaid on the K-band image from VLT/NACO are X-ray (blue) contours at 2, 4, 8, 16, and 32 $\times 10^{-4} {\rm~cts~s^{-1}~arcsec^{-2}}$, which are obtained from the \chandra/ACIS-S/HETG 0th-order intensity image constructed in the 1-9~keV range and adaptively smoothed with {\it CSMOOTH} with a signal-to-noise ratio of 2.5-3.5. (B) An unsmoothed X-ray image of the region from later ACIS-S observations with no grating; the brightest source, 2\farcs4 southeast of Sgr A*, in the image is the magnetar SGR J1745-2900, which was not active during the period when the data in (A) were taken~\citep{mori13}. The spectral extraction regions for \xs\ and its background are outlined by the circle of 1\as\ radius (colored in cyan) and the asymmetric annuli (magenta) around \xs. North is up and East is to the left for all the presented images throughout the paper.  
}
\label{f:f1}
\end{figure*}

  Here we focus on \xs, which is located only 3\farcs6  (1''=0.039 pc at our adopted GC distance $D=8$~kpc) in projection away from Sgr A*, compared with its Bondi radius of 4'' for hot gas accretion~\citep{wang13}. 
  This apparent proximity to Sgr A* makes the group particularly interesting and puzzling~(e.g., Fig.~\ref{f:f1}A). Within a radius of 0\farcs27, the group contains two Wolf-Rayet (WR) stars (E2 -- WN8 and E4 or more precisely E4.0 -- WC9) and one OB supergiant (E1), as well as numerous red objects, including E3, which has been resolved into no less than six stable components \citep[e.g.,][]{fritz10}. The three confirmed massive stars, sharing similar proper motions and line-of-sight (LoS) velocities (Table~\ref{t:IRS13E-para}), appear to be gravitationally bound. 
But the total stellar mass of the stars ($\sim 350{\rm~M_\odot}$ according to \citealt{paumard06} or  up to $\sim 2000{\rm~M_\odot}$  to \citealt{fritz10}) is apparently far too little to overcome the strong tidal force of Sgr A*, if the projected separation  approximately represents the physical one. The required self-gravitating mass indicates the presence of an embedded intermediate-mass black hole (IMBH), with a mass $M_{IMBH} \gtrsim 10^3 {\rm~M_\odot}$~\citep{maillard04} and possibly as large as several $10^4 {\rm~M_\odot}$~\citep{schodel05}, and/or many stellar-mass ones~\citep{fritz10,banerjee11}. This upper bound mass is obtained from the existing limits on the proper motion acceleration of stars in \xs\ and on the   reflex motion of Sgr A*~\citep{fritz10}. 

Interestingly, an IMBH of $2.4 \times 10^4~{\rm M}_{\sun}$ has recently been proposed to be responsible for the large velocity width ($\sim 650 {\rm~km~s^{-1}}$) of the spatially resolved ionized gas detected at \xs~3.0~\citep{tsuboi17,tsuboi19}. Whether or not such an IMBH is present in the group has strong implications for understanding the gas and stellar dynamics in the GC, as
well as \xs\ itself~\citep{portegies-zwart06}. We find that the large velocity width, as well as some other outstanding features in and around \xs, may alternatively and naturally be explained by the collision of the fast stellar winds from the group with its ambient medium. The group wind can then be partially confined by the ram-pressure of the collision, while dense blobs are compressed, evaporated and stripped, forming gas flows of high density and velocity spread as observed.  Furthermore, the dispersed cool gas, as well as the \xs\ wind, can be an important source of mass and/or ram-pressure in the vicinity of Sgr~A*~\citep{melia01,paumard01,Moscibrodzka06,cuadra08}, affecting the gas dynamics around the SMBH and even its accretion. Here we attempt to address these issues about the nature of \xs\ and its interplay with the environment, based on hydrodynamic simulations and multi-wavelength observations. 

\begin{table}
\begin{center}
\caption{Velocities of objects in \xs}
\begin{tabular}{lcccr}
\hline\hline
object      &    $v_{\rm RA}$ &$v_{\rm Decl.}$ &   $v_z$     & $v_{\rm T}$\\
E1           &  -142.5  &   -105.5     & 134$\pm10$ &222\\
E2           &  -249.2  &       23     & 65 $\pm30$ &259\\
E3.0        &   -82     &  8           &  -23 &86\\
E4        &  -227     &  26          & 200$\pm200$ &304\\
\hline
$\bar{v}_{E2+4}$  & -238      & 25          & $68\pm30$ & 249\\  
$v_{bar}$      & -150&    64        & -30       &166\\
$v_r=\bar{v}_{E2+4}-v_{bar}$ & -88 & -39& 98 &  137\\ 
\hline
\end{tabular}
\end{center}

Note: The velocities are  in units of ${\rm~km~s^{-1}}$. The 
measurements for E1-4 are from Table 1 and Appendix B of~\citet{fritz10}.
Quoted errors for the transverse components $v_{\rm RA}$ and $v_{\rm Decl.}$ are not included because they are negligible compared to other uncertainties; 
 $v_{\rm T}$ is the derived total velocity of each object. The average velocity  ($\bar{v}_{E2+4}$) of the two energetically dominant stars, E2 and E4, is assumed as the effective velocity of \xs.  While the LoS component ($v_z$) of the bar
 velocity ($v_{bar}$) is estimated with the \alma\ data presented here (see \S~\ref{ss:alma-results}, in particular Figs.~\ref{f:p-v-diag} and \ref{f:v-profs}), the $v_{\rm RA}$ and $v_{\rm Decl.}$ components are with the mean {\sl VLA}-measured proper motion of gas local to \xs\ (see \S~\ref{ss:dis-intr}), together with the  position angle of 25$^{\circ}$ of the bar~\citep{liszt03}.
\label{t:IRS13E-para}
\end{table}

We start with an analysis of \chandra\ observations on \xs. The group is a known X-ray source \citep{baganoff01}. \citet{wang06} have shown that its emission is extended in a \chandra\ image and that its X-ray spectrum is hard, inconsistent with what is expected from individual massive stars, but typical for colliding-wind (CW) binaries~\citep{coker00,coker02}. However, many questions remain unanswered: What may be the morphology of the extended emission, if confirmed?  What is the true nature of the emission?  Is there any evidence for an accreting BH? Could it  really be produced by the stars well separated spatially in \xs? What can we learn about the interplay between \xs\ and its environment from the properties of the X-ray emission? To address such questions, we conduct a systematic spatial and spectral analysis of the X-ray emission from \xs, based on the archival \chandra\ observations with a total of 5.3 Ms exposure time, as reported here. We further perform hydrodynamic simulations to check the consistency of the interpretation of the X-ray emission as being due to the CW~\citep{coker00,coker02}. In the process, we also provide useful constraints on relevant parameters of \xs.   

To explore the interplay between \xs\ and its environment, we also analyze a set of \alma\ observations, as well as relevant results from the literature~\citep[e.g.,][]{liszt03,zhao09,fritz10,eckart13,tsuboi17,tsuboi17a,tsuboi19}. This analysis allows us to better assess the quality of the \alma\ data and to use them in examining various hypotheses, as well as testing our group wind collision scenario. 

The organization of the paper is as follows: After a brief description of the X-ray data and their reduction in \S~2, we present our detailed analysis and results in \S~3, discuss constraints on the X-ray emission from the putative IMBH in \S~4, and compare our observational results with the simulations in \S~5. We describe the \alma\ data analysis and results in \S~6. We further present a critical analysis of existing multi-wavelength results and interpretations in \S~7, before moving on to test our group wind collision scenario and to explore its implications in \S~8. Finally in \S~9, we summarize our results and conclusions. 

\section{\chandra\ Observations and Data Reduction}\label{s:obs}

The \chandra\ observations used in the present work can be divided
into three sets. The first set, consisting of 45 separate observations (1.5 Ms total exposure) taken before 2012, had the Advanced CCD Imaging Spectrometer-Imaging (ACIS-I) on axis. The second set of 38 observations (3 Ms total) taken during February 6 to October 29, 2012 were from the  X-ray
Visionary Project (XVP) of Sgr A*, which utilized the High
Energy Transmission Gratings (HETG) in combination with the Advanced
CCD Imaging Spectrometer-Spectroscopy (ACIS-S; only the 0th-order
ACIS-S/HETG data are used here). The third set, including 25 observations with the ACIS-S without the grating, were taken after the XVP, but before May 15, 2015. 
The combination of these observations, pointed within  2' of Sgr A*, greatly enhances the counting statistics for the present exploration. The different sets of the data are complementary in their
qualities. The spectral resolution of the ACIS-S data is about a factor of 
$\sim 2$ better than that of the ACIS-I observations (which suffered from a severe charge transfer inefficiency and cannot be fully mitigated in software processing). The 0th-order data of the ACIS-S/HETG observations have the longest
total exposure and best counting statistics, although the effective collecting area is low compared to the ACIS-S/NONE (without the grating) data, especially at lower energies ($\lesssim 4$~keV). 
The ACIS-I data, with the effective area considerably higher than that of the 
ACIS-S at higher energies, provide the most sensitive coverage of 
the spectrum above the 6.7-keV iron line.

Individual observations in each of the three data sets are reduced via standard {\small CIAO} processing routines. 
We remove time intervals of significant background flares, with count rates deviating more than 3$\sigma$ or a factor of $\gtrsim 1.2$ from the mean rate 
of individual observations, using the light-curve cleaning routine \textsc{LC$\_$CLEAN}. This cleaning, together with a correction for the dead time of the observations, resulted in 1.44 (ACIS-I), 2.92 (ACIS-S/HETG) and 0.82 (ACIS-S/NONE) Ms effective exposures for 
subsequent analysis. We correct for the relative astrometry of the observations by matching 
the centroid positions of sources detected within $3^{\prime}$ of 
Sgr A*, 
and then align the absolute position to the radio position of Sgr A* [J2000: 
$RA=17^h45^m40\fs041, Dec=-29^\circ 00^\prime 28\farcs12$;~\cite{reid04}]. 
The three data sets with significantly different instrument responses are analyzed separately and allow independent reliability tests of our various measurements. The source detection is the same as detailed in the supplementary material of \cite{wang13}.

To study the spatial structure of \xs, we need to account for both the point-spread function (PSF) of the instrument and the scattering of the X-ray emission by dust along the line of sight. To do so, we construct an empirical spatial  distribution of the X-ray intensity from J174538.05-290022.3, which shows high time-variability and is thus point-like. With a spectrum indicative of strong
absorption and projected only 24$^{\prime\prime}$ west of \xs\ [see the supplementary material of~\cite{wang13}], 
this source should be located at the GC or beyond. The source has a count rate of $5.6 {\rm~counts~ks^{-1}}$ in the ACIS-S/HETG 0th-order image, indicating a slight pile-up effect [two or more photons in a single ``readout frame''  being registered as a single event, or rejected as a non X-ray  event;~\citet{davis01}]. The pile-up makes the intensity distribution at
the centroid of the source a bit flat, but should have negligible effect on the overall distribution. The count rate of the source  is much higher in the ACIS-I or ACIS-S/NONE image. The pile-up effect is correspondingly more important, making the source not an appropriate PSF calibrator. Therefore,  our spatial analysis is limited to the ACIS-S/HETG (0th-order) data.

\section{X-ray Data Analysis and Results}\label{s:res}

Fig.~\ref{f:f1} presents the X-ray intensity images from the ACIS-S/HETG (A) and ACIS-S/NONE (B)
observations. In addition to Sgr A* and \xs, there are two discrete X-ray sources that deserve some notes. First, the bright source seen in Panel (B), 2.4'' southeast (SE) of Sgr A*, is the magnetar, SGR J1745-2900, which appeared only
after 2012~\citep{mori13}.  Second, the X-ray enhancement at E60~\citep{paumard06}, just west of \xs\ (Fig.~\ref{f:f1}A),  has a hard spectral shape, as indicated by the 4-9~keV to 1-4~keV band intensity ratio, and is apparently due to the CW of this eclipsing binary.  It has a short orbital period of 2.28 days and an inclination angle of $\sim 70^{\circ}$, as determined from both spectroscopical and photometric observations~\citep{pfuhl14}. These observations further suggest that both the primary ($\sim 20$~M$_{\odot}$) and the secondary (10~M$_{\odot}$) of the binary have similar spectral types (WN7).  The binary is unlikely to be bound to \xs.  Although the proper motion of E60~\citep[object \#376 in][]{schodel09}  is consistent with that of \xs, their LoS velocities are substantially different, $\sim 467 {\rm~km~s^{-1}}$ (E60) vs. $\sim 60 {\rm~km~s^{-1}}$ (IRS13E), indicating that the binary may even be not bound to the gravity of Sgr A*~\citep{pfuhl14}. 

\subsection{Spatial analysis}
\label{ss:res_spat}

As shown in Fig.~\ref{f:f1}A, the X-ray emission from \xs\ appears extended and peaks in the middle between the two WR stars, E2 and E4. To be more quantitative, we compare the centroid position of the X-ray peak with those of the two WR stars. We first measure the positions of the stars in the IR image (taken in 2005) and then correct them according to the known proper motions~\citep{paumard06}, which are about 0\farcs05 over a seven year period.
We find that the separation of the two stars is 0\farcs32, mostly in the N-S direction. The errors in these measurements, as well as the X-ray one,  are difficult to quantify exactly and are mostly systematic, but should be smaller than $\sim 0\farcs1$.  The X-ray centroid is right in the middle (slightly closer to E4) in the N-S direction, but is about 0\farcs15 off to the west.

\begin{figure*}https://www.overleaf.com/project/5d40d06389e5832705b14a32
\centering
\includegraphics[width=0.5\linewidth,angle=270]{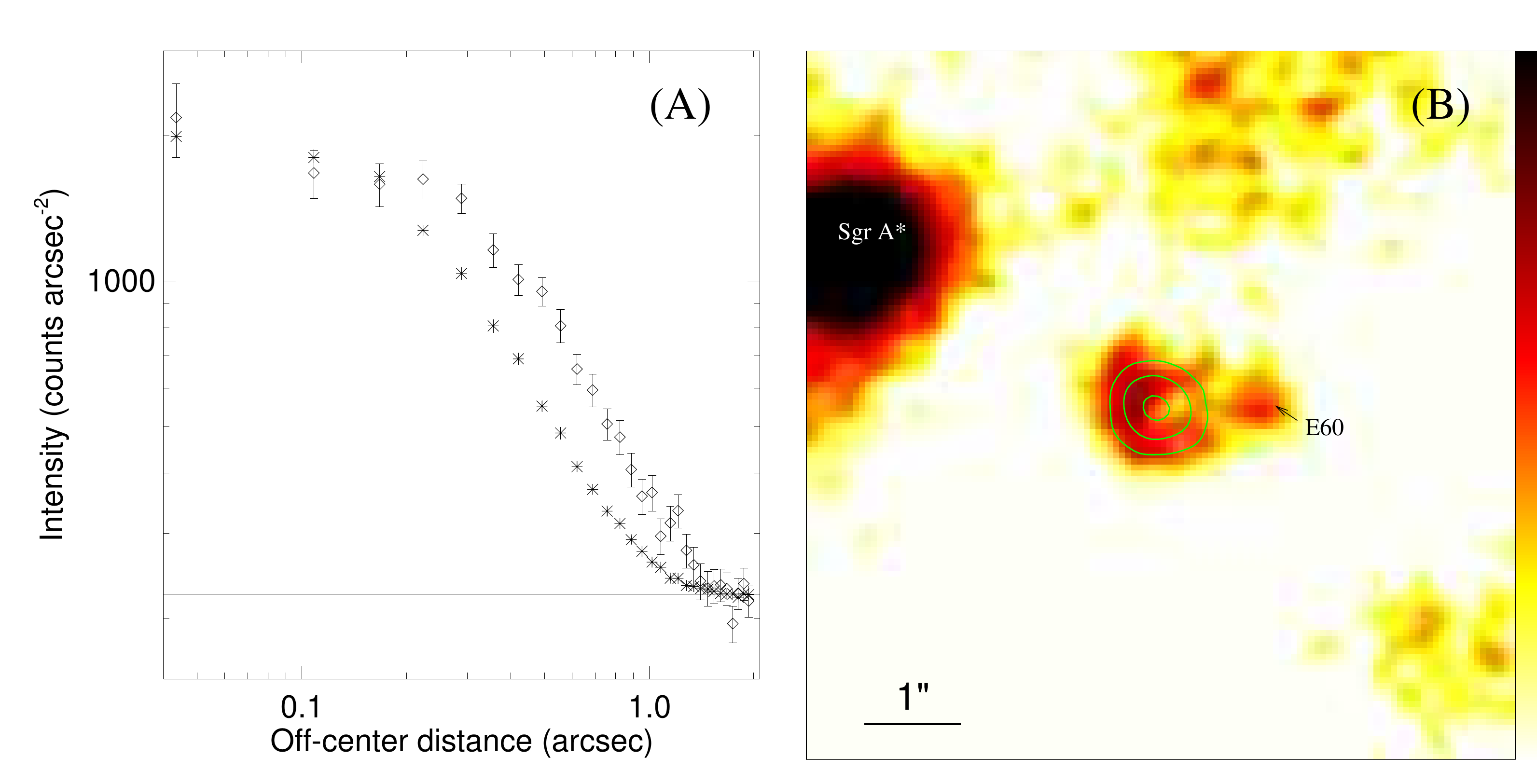}
\caption{
(A) The radial  intensity profile of \xs\  ({\sl diamonds}), compared with the profile of the point-like source J174538.05-290022.3 ({\sl crosses}), all in the ACIS-S/HETG 1-9~keV band. The latter profile has been normalized to the average intensity of the first three bins of the former. This normalization is performed after subtracting the corresponding local background of each profile, estimated as the median value in the 1\farcs5-2$^{\prime\prime}$ range. The profile is then shifted to the local background level (the solid horizontal line) of the \xs\ profile. (B) A close-up of the ACIS-S/HETG 1-9~keV intensity image of \xs. The intensity is decomposed into extended (color image) and point-like (contour) components.  The color bar of the former component is scaled logarithmically in the 0.3-1.5 $\times 10^3$ counts~arcsec$^{-2}$ range (white-black), while the latter component is modeled with the emission from the point-like source,  J174538.05-290022.3, and is illustrated as the intensity contours at 0.3, 0.5, 1.0, and 2 $\times 10^3$ counts~arcsec$^{-2}$.
}
\label{f:f2}
\end{figure*}

The extent of the X-ray emission can be appreciated better in Fig.~\ref{f:f2}A, which compares a radial X-ray intensity profile of \xs\ with that of the point-like source J174538.05-290022.3. To obtain an upper limit to a potential point source contribution to the emission, we subtract from its image a normalized image of J174538.05-290022.3 (Fig.~\ref{f:f2}B). This latter image is scaled to minimize the central peak of the resultant image, but not to produce a hole because of the subtraction.
The adopted scaling factor corresponds to a subtracted point-like component
accounting for 50\% of the total  emission. This component can then be considered to be the maximum contribution of a central point source, if present. The remaining diffuse emission in the immediate vicinity of \xs\ has the full size of about 2$^{\prime\prime}$ with an uncertainty of $\sim 20\%$, accounting for the PSF/dust broadening (see \S~\ref{s:simul} for more discussion).

The morphology of the X-ray emission is also interesting. The morphology seen in Fig.~\ref{f:f2}B, however, somewhat depends on how the ``point-like'' component is removed, particularly the choice of its centroid position. Nevertheless, the emission is qualitatively most extended in the N-S direction on the east side of \xs, which is apparent also in Fig.~\ref{f:f1}A. The extendness in the E-W direction is also apparent on the east side. The extendness to the west side is uncertain, largely because of the confusion with E60. We see no evidence for physical interaction between this binary and \xs. Overall, the morphology of the emission seems to fan out toward the east,  roughly in the opposite direction of the proper motion of \xs~\citep[][ and see \S~\ref{s:simul} for more discussion]{paumard06}.
https://www.overleaf.com/project/5d40d06389e5832705b14a32

\subsection{Spectral analysis}
\label{ss:res_spec}
\begin{figure}
\unitlength1.0cm
\centerline{
\includegraphics[width=1\linewidth,angle=0]{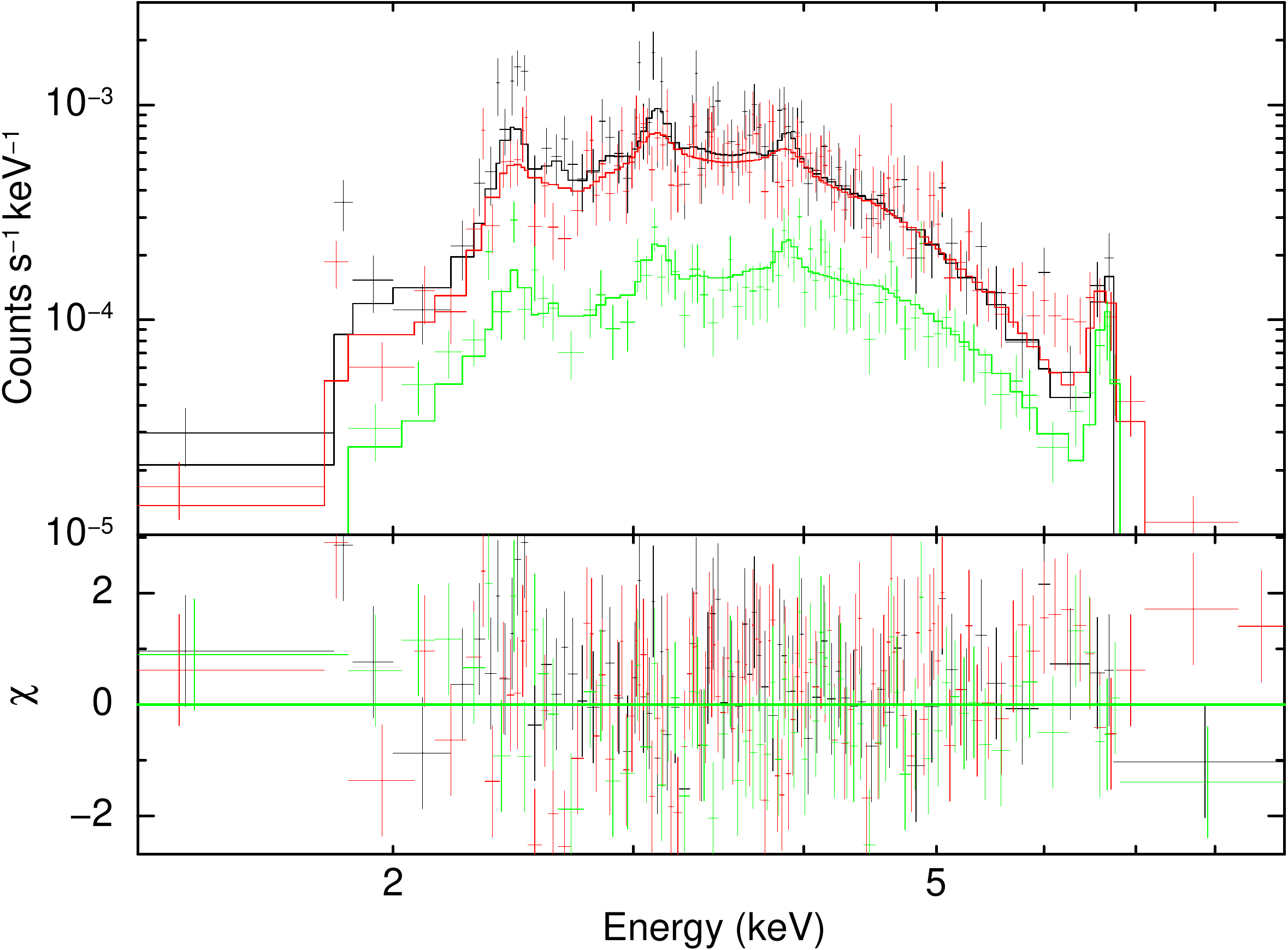}
}
\caption{\chandra\ spectra of \xs, together with the best-fit XSPEC {\it VAPEC} model (Table~\ref{t:para}): ACIS-S/NONE (black) and ACIS-S/HETG  0th-order (green) and ACIS-I (red).
}
\label{f:f3}
\end{figure}

Fig.~\ref{f:f3} presents X-ray spectra of \xs, extracted from the three data sets. These spectra are local background-subtracted (see the spectral extraction regions shown in Fig.~\ref{f:f1}B) and are adaptively grouped to achieve a signal-to-noise ratio greater than 3 per bin. The presence of the S, Ar, and Fe He-$\alpha$  emission lines is apparent, clearly indicating the thermal nature of the spectra.
We detect no significant variation in the flux and spectral shape among the spectra, so we fit them jointly, first with an optically thin plasma {\it VAPEC}. (Here the spectral models are all  in reference to those implemented in the XSPEC software package; \citealt{arnaud96}.) 
The photoelectric absorption uses the Tuebingen-Boulder model [{\it TBABS}; \cite{wilms00}] with the cross-sections from \cite{verner96}. The metal abundance pattern of the plasma and the foreground
X-ray-absorbing gas are adopted from~\citet{wilms00} for the interstellar medium (ISM). 
This simple model fit is acceptable and gives the best-fit parameters listed in Table~\ref{t:para}. In particular, the fitted X-ray absorbing, equivalent hydrogen column density $1.74 (1.64-1.94) \times 10^{23}\rm~cm^{-2}$ for \xs, which may be compared with $1.66 (1.41 - 1.94) \times 10^{23}\rm~cm^{-2}$ for Sgr A*~\citep{wang13}. 

\begin{table}
\begin{minipage}[t]{3in}
\caption{X-ray characterization of \xs}
\begin{center}
\begin{tabular}{@{}lr}
\hline\hline
Centroid position &$17^h45^m39\fs78$\\
& $-29^\circ0^{\prime}29\farcs7$\\
Overall size & $\sim 2^{\prime\prime}$\\
Plasma temperature kT (keV) & 1.34 (1.22-1.46)\\
ISM metal abundance$^a$ &1.1(0.83-1.7)\\
Absorbing column density N$_{rm H}$ ($\times 10^{23}\rm~cm^{-2}$) & 1.74 (1.64-1.94)\\
Normalization ($10^{-4}$)& 5.0(4.4-6.9)\\
$L_{1-9\rm~keV} ({\rm erg~s^{-1}})$ &1.2$\times$10$^{33}$ \\
$\chi^2/d.o.f.$ & 314/267\\
\hline
\end{tabular}
\end{center}

  Note: The metal abundances are relative to the interstellar medium (ISM) values given in~\citet{wilms00}. The presented error ranges are at the 90\% confidence.
\label{t:para}
\end{minipage}
 \end{table}

We examine how various modifications of the plasma  model may affect the estimation of the spectral parameters. 
For example, one may expect that the ionization state of the plasma takes time to establish an equilibrium.  We thus fit the spectra with the XSPEC {\it VNEI} model, even with separate grouping of metals into $\alpha$-elements and Fe-like elements. We find that this model does not significantly improve the fit to the spectra (e.g., $\chi^2/d.o.f. =$  310/265). The modeling does give a probably meaningful constraint on the ionization parameter as  $\tau= 2.3(1.2-5.2) \times 10^{11} {\rm~s~cm^{-3}}$. But other model parameters are hardly changed, compared to those listed in Table~\ref{t:para}.  

We further consider the possibility that the metal abundance pattern of the X-ray-emitting plasma in \xs\ could be substantially different from that of the ISM. The stellar wind from the WC star, in particular, is rich in Carbon, presumably due to the early triple-$\alpha$ nuclear synthesis  and exposed now due to the mass loss.  Although the X-ray emission lines of Carbon ions are too low energies to show up in the observed spectra of \xs, because of the strong LoS X-ray absorption, the Bremsstrahlung continuum can be enhanced due to its proportionality to $Z^2$, where $Z$ is the charge of an ion. (It is easy to show that H and He nuclei have the same efficiency, per unit mass, for producing the Bremsstrahlung continuum.) This continuum enhancement due to the enrichment of a particular metal, if not specifically  considered as in the above model fits, could in general lead to an underestimate of the metal abundance.  For a WC star, we may assume that only the Carbon enrichment is important, while the abundances of other tracer metals  (e.g., Fe) remain unchanged, compared to those in the ISM. The continuum enhancement factor is then $ [Zf_{Z}/2+(1-f_Z)]$, where $Z=6$ is the atomic number of Carbon (the factor of 2 arises from the ion mass $2Z$) and $f_{Z}$ is the mass fraction of the ion.  Because $f_{Z}$ can be up to $\approx 0.5$ (Koesterke \& Hamann 1995), this factor can reach to $\sim 2$.  The resultant underestimate of the metal abundance can then be up to the same factor.  In contrast, the Nitrogen enrichment of a WN star should have little effect on the Bremsstrahlung efficiency. This enrichment is due to the conversion of C and/or O to N in the CNO cycles of hydrogen burning. Thus the stellar wind of a WN star is primarily composed of Helium. Therefore, with a more realistic stellar wind contributions from the WC and WN stars in \xs, the metal abundance underestimate should be quite modest. 
With this estimate in consideration, we conclude that our metal abundance listed in Table~\ref{t:para} is consistent with the previous measurements~\citep{cunha07,davies09,wang06,capelli12}. 

We estimate the mean electron density of the plasma in \xs\ as $n_{e}\sim 6 \times 10^{2} {\rm~cm^{-3}}$, based on the size of the diffuse X-ray emission region, as well as the normalization listed in Table~\ref{t:para} and its standard XSPEC definition. Together with the plasma temperature, this leads to an estimate of the pressure as $P/k \sim 2 \times 10^{10} {\rm~K~cm^{-3}}$. The plasma is thus strongly over-pressured, compared to the general ISM in the GC ~\citep[$P_{ISM}/k\sim 10^{5} -10^{9} {\rm~K~cm^{-3}}$; e.g.,][]{rathborne14}.

\section{Constraints on X-ray emission from the putative IMBH}

We check whether an IMBH in \xs\ may contribute to the observed X-ray emission. 
In \S~\ref{ss:res_spat}, we have used the spatial decomposition of the observed X-ray emission to constrain the contribution from a potential central point-like source in \xs. This source could represent the accretion of the stellar wind material by this putative IMBH. Our upper limit on the source luminosity, $L_x \sim 6 \times 10^{32} {\rm~ergs~s^{-1}}$ in the 2-10 keV band (\S~\ref{ss:res_spat}), is certainly far less than the Eddington limit of the IMBH, $L_{Edd} \sim 2 \times 10^{42} M_{IMBH,4} {\rm~ergs~s^{-1}}$, where  $M_{IMBH,4}$  is the IMBH mass in units of $10^4{\rm~M_\odot}$. 
The IMBH, if present, should then be undergoing a radiatively inefficient accretion, as in the case for  Sgr A* (e.g., ~\citealt{wang13}).

Could such a radiatively inefficient IMBH explain the possible point-like component of the X-ray emission? In the hot accretion case, the spatial distributions of the plasma density and temperature of the accretion flow can be scaled with their values at the so-called Bondi radius $r_{b} \propto M_{IMBH}$~\citep[e.g.,][]{roberts17}. This radius characterizes the size of the region in which the plasma of certain thermal and/or kinematic properties can be initially captured by  the gravity of the black hole. Although the Bondi accretion model itself is  typically too simplistic to describe the complex accretion process, the X-ray emission from the hot gas accretion is qualitatively dominated by the outer accretion flow~\citep[e.g.,][]{wang13,roberts17}. 
One may then expect that $L_x \propto r_{b}^{3} n_e^{2}  \propto M^{3}_{IMBH} n_e^{2}$. 
Our estimated $n_{e}$ in \xs\ is a factor of nearly 4 higher than the value at the Bondi radius of Sgr A*~\citep{roberts17}. Since Sgr A* is $L_{x} \approx 3.4 \times 10^{33} {\rm~ergs~s^{-1}}$, the luminosity of the accreting IMBH should then be $\sim 9 \times 10^{26}  M^3_{IMBH,4}  {\rm~ergs~s^{-1}}$, far too weak to explain the X-ray emission of \xs. 

The IMBH might, however, be located very close to one of the WR stars, and be accreting from its freely expanding stellar wind. In this case, the wind density decreases with the distance $r$ from the star. Our upper limit to the point-source luminosity, for example, would be reached if $r \lesssim 6\times 10^{-4} {\rm~pc}~\dot{M}_{w,-5}^{1/2} v_{w,3}^{-1/2}$, where $\dot{M}_{w,-5}$ 
and $v_{w,3}$  are the mass-loss rate and stellar wind speed in units of $10^{-5} {\rm~M_\odot~yr^{-1}}$ and $10^{3} {\rm~km~s^{-1}}$, respectively.  This distance corresponds to $\lesssim 0.1^{\prime\prime}$ in projection, which means that the X-ray source should  spatially almost coincide with one of the WR stars, which is inconsistent with our measurement of the X-ray peak centroid (\S~\ref{s:res}).  Therefore, the putative IMBH, accreting from the  stellar wind material, is not expected to contribute significantly to the observed X-ray emission. 

\section{Hydrodynamic simulations and comparison with the X-ray observation}
\label{s:simul}

The spatial and spectral properties of IRS~13E  strongly suggest that its X-ray emission arises primarily from the CW in the group~\citep{coker02}. In this scenario, the stellar winds are heated in standing shocks around the stars. The characteristic temperature of the shock-heated gas is $1.5 \times 10^{7}  v_{w,3}^{2}$~K, roughly consistent with the measured value in \xs\ (Table~\ref{t:para}). At least part of the X-ray emission from \xs\ must originate in the CW. The mechanical energy input rate from the stars is $L_{\xs} \sim (3 \times 10^{37} {\rm~ergs~s^{-1}}) \dot{M}_{-4} v^{2}_{w,3}$.  Only $\sim 10^{-4}$ of this wind energy input is required to produce the  X-ray luminosity of \xs. The rest of the energy must be carried out by the combined group wind, escaping into a larger region and regulating the environment of the Galactic nucleus \citep[e.g.,][]{calderon20}.

Here we attempt to check the consistency of the CW scenario with the above observational results more quantitatively, via simulations. We first update hydrodynamic simulations of the central parsec around Sgr~A* that include IRS~13E and then compare synthesized spectra and images with the observed ones.  This part of our study follows the same method of \citet{russell17}, the relevant details of which we recap here. 

With the purpose of studying the gas dynamics around Sgr~A*, \citet{cuadra08, cuadra15} performed smoothed particle hydrodynamic simulations of the central 25 WR stars and their stellar winds within 1~pc of the SMBH.  Using the location of the stars 1100 yr ago as the initial condition, the simulations had the stars orbit the SMBH while ejecting gas particles to model their stellar winds up to a few hundred years into the future.  The construction of the orbits used the existing 3-D velocity measurements of individual stars and their positions on the sky, as well as their assumed orbit shape (quasi-circular or following the so-called clockwise disk \citep{paumard06}), which constrain their unknown LoS locations. But this constraint can be rather uncertain for specific objects such as stars in the IRS~13E group. The orbits of individual stars within the group are also unknown, depending largely on its assumed total mass. The \citeauthor{cuadra08} simulations included an IMBH of 350 M$_\odot$ in between the stars of the group~\citep{fritz10}.  The winds from the 25 WRs naturally collide with one another, creating a complex, time-varying reservoir of shock-heated gas, a small fraction of which is captured by Sgr~A*.  

The hydrodynamic simulations, confronted with \chandra\ observations, can be used to test the CW scenario, and to constrain the relative line-of-sight position and the wind properties of the stars.  \citet{russell17} synthesized the spatially resolved thermal X-ray emission from the simulations by using the optically-thin plasma {\it VVAPEC} model  for the emission, which was split into three abundance-specific spectral classes: WC, early WN, and late WN.  The resulting pixel-by-pixel spectra were folded through the instrument response function, and then 4-9 keV images and 2\arcsec-5\arcsec annulus spectra were extracted to compare directly with the observations.  The best-matched model, which incorporated feedback from the SMBH to clear out some of the hot gas from the region surrounding Sgr~A* \citep{cuadra15}, agreed in the image and spectra to within ~20\%.  However, the IRS~13E group was not well matched; the synthesized X-ray emission is greater than the observed flux, regardless of Sgr A* feedback model, indicating either that the stars were too close in 3D separation and/or that the wind parameters used, i.e.\ mass-loss rates and/or wind speeds, were too high (see below). 

We conduct multiple simulations tuned to match the \chandra\ data of \xs. Its X-ray spectral shape in the adiabatic CW scenario considered here depends primarily on the stellar wind speeds of individual stars, chiefly E2 and E4, but also on their {\sl relative} mass-loss rates. The spatial extent of the X-ray emission scales linearly with the separation between two stars ($\delta D$), which cannot be directly observed and could be much greater than what is assumed in the simulation (roughly the projected distance of $\sim 0\farcs3$). The luminosity of the emission scales with the combined mass-loss rate of the two stars ($\dot{M}_w$), as well as $\delta D$, according to
\begin{equation}
L_x \propto \dot{M}_w^2/\delta D.
\label{e:norm}
\end{equation}
Our simulations sample various stellar wind speed and mass-loss rate ratios, around the predictions from the stellar types of the stars \citep[][ and references therein]{russell17}, while the remaining 23 WR stars in the simulations are unchanged. From the comparison of the simulated X-ray spectra and sizes with the observed ones of \xs, we constrain $\delta D$, as well as the stellar wind speeds and mass-loss rates of the stars. 

\begin{figure}
\centering
\includegraphics[width=1\linewidth]{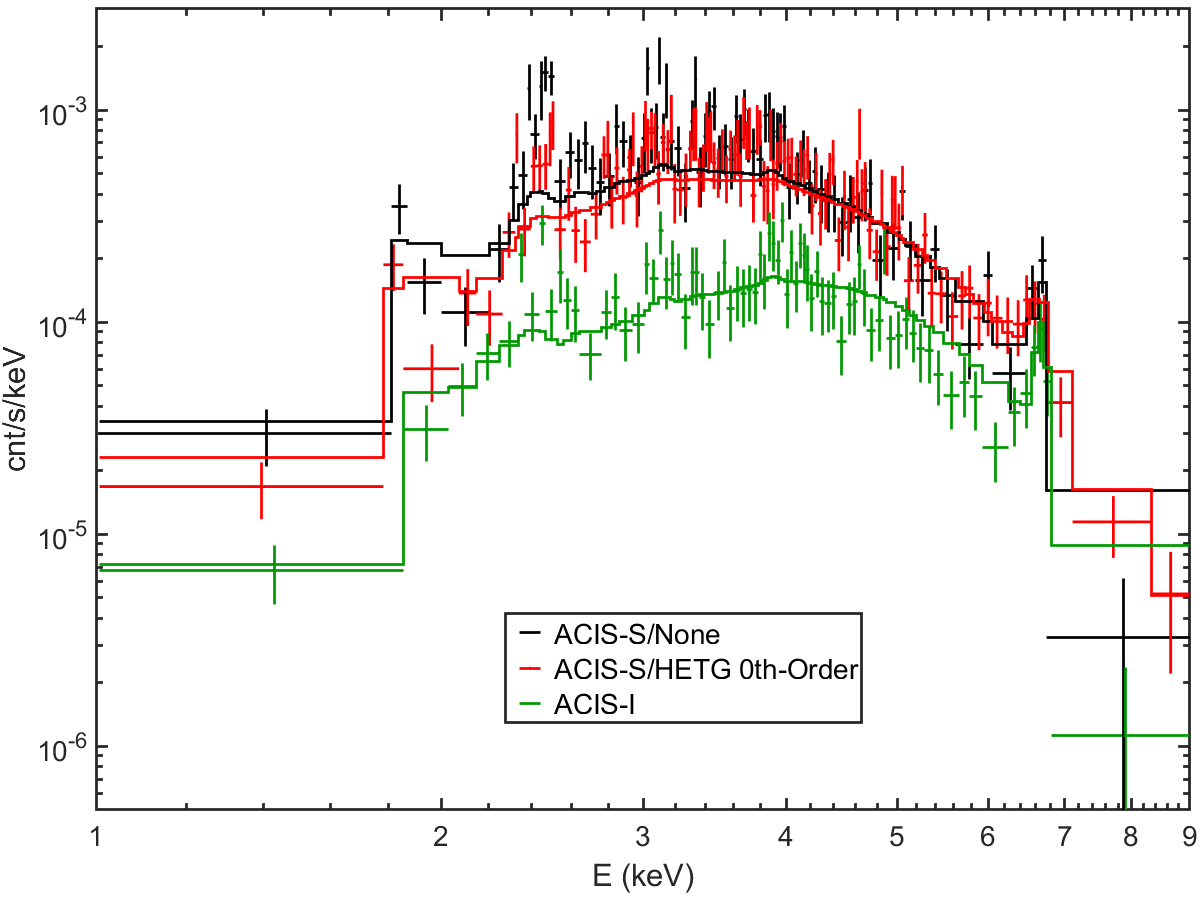}
\includegraphics[width=1\linewidth]{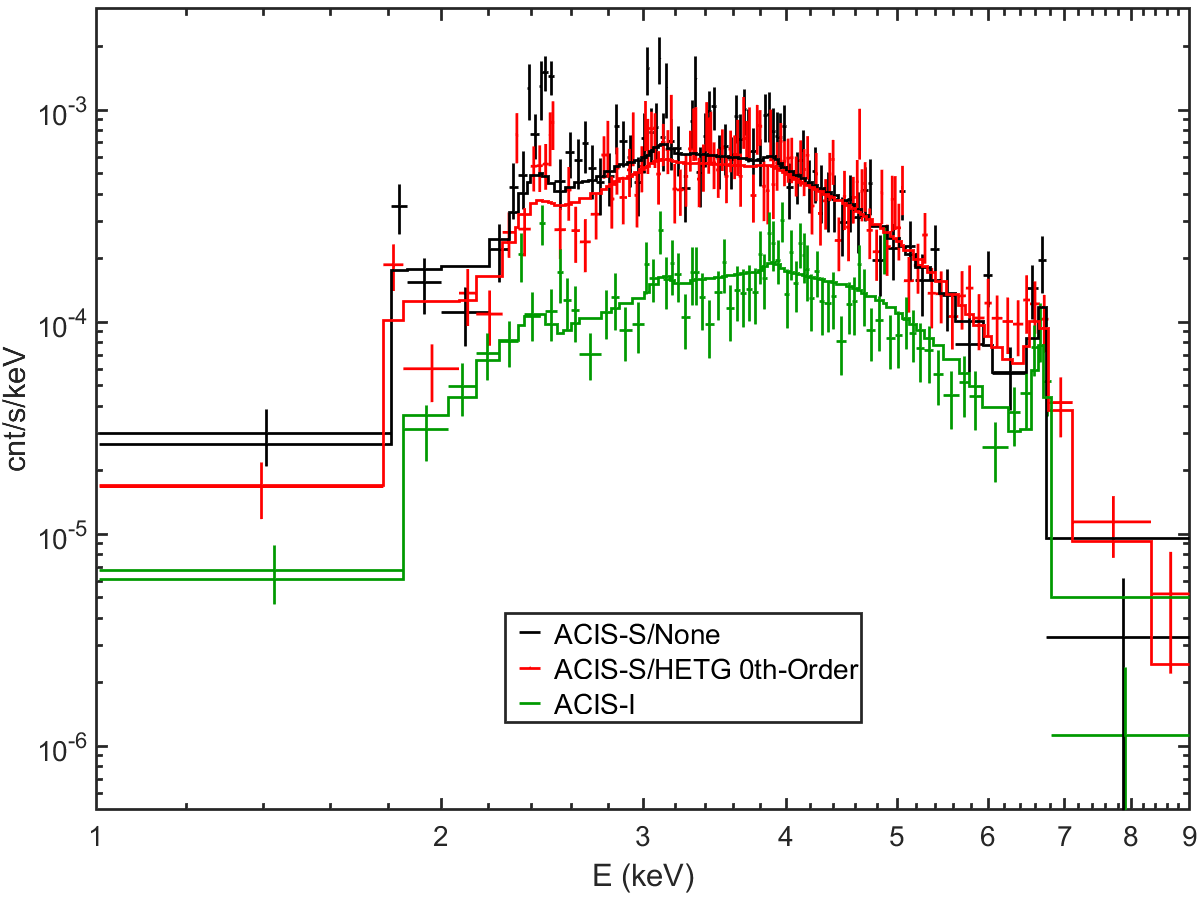}
\caption{Representative fits of simulated model spectra to the same \chandra\ data as in Fig.~\ref{f:f3}: our initial "best guessed" model (upper panel) and the "best fit" model (lower panel) with the parameters listed in Table~\ref{t:WindParam}. 
}
\label{f:spec_sim}
\end{figure}

For the new set of simulations run for this work, we start with a "best guessed" set of stellar wind parameters for the two WR stars in \xs. We reduce the WC wind as its analysis was subject to significant dust contamination, which might be the reason for its wind speed and mass-loss rate apparently being larger than the other WC stars in the GC~\citep{martins07}.  For the WN wind, updated modeling of the mass-loss rate in other GC WN stars reduces the mass-loss rate by a factor of $\sim$2~\citep{yusef-zadeh15}.  The reduction factors of 2 for both $\dot{M}$ and 1.5 for the WC wind speed provides our initial "best guessed" model presented in Table~\ref{t:WindParam}.
In the subsequent runs, we adjust the parameters one by one, not only to bring the model into better agreement with the model, but also to verify expected trends in how the X-ray emission is sensitive to the wind speeds and the wind momentum ratio. 

To have as detailed a model-to-observation comparison as possible, we process the model images and spectra in the same manners as for the observations.   Specifically, the spectral fits are performed in an analogous method to the \textit{VVAPEC} fitting of the observed data, namely using \textsc{xspec} to simultaneously fit all three \textit{Chandra} spectra where the hydrodynamic+emission model spectra, extracted in the same projected regions, are absorbed by the ISM according to the \textsc{tbabs} model with \textsc{wilms} abundances.  This absorption is fixed at the best-fit ${\rm N_H}$ (Table~\ref{t:para}) to reduce the number of variables for the present modeling.

\begin{table*}
{\center
\caption{Model parameters of the \xs\ WR stars}\label{t:WindParam}
\begin{tabular}{lrrr}
\hline\hline  
Parameter & \multicolumn{3}{c}{Value}\\
  & ``best guessed'' & ``best fitted''& `normalized best fitted''\\  
\hline
E2 (WC) $\dot{M}$ & $2.5\times10^{-5} {\rm~M_\odot~yr^{-1}}$ &$3.3 \times 10^{-5}  {\rm~M_\odot~yr^{-1}}$ &$6.3 \times 10^{-5}  {\rm~M_\odot~yr^{-1}}$\\
E2 (WC) $v_{w}$ & 1500 ${\rm~km~s^{-1}}$& 750 ${\rm~km~s^{-1}}$ & 750 ${\rm~km~s^{-1}}$\\
E4 (WN) $\dot{M}$& $2.2 \times 10^{-5}  {\rm~M_\odot~yr^{-1}}$ & $2.9 \times 10^{-5}  {\rm~M_\odot~yr^{-1}}$ & $5.5 \times 10^{-5}  {\rm~M_\odot~yr^{-1}}$\\
E4 (WN) $v_{w}$& 750 ${\rm~km~s^{-1}}$  & 750 ${\rm~km~s^{-1}}$ & 750 ${\rm~km~s^{-1}}$\\
$\delta D$ & 0.013~pc ($\equiv 0\farcs34$) & 0.014 ($\equiv 0\farcs34$) & 0.05~pc ($\equiv 1\farcs3$)\\
\hline
\end{tabular}
}

Note:  The initial $\dot{M}$ values for the "best guessed" model given here are not adjusted for a small normalization correction that gives the fit to the \chandra\ spectra shown in Fig.~\ref{f:spec_sim} upper panel. The "best fit" model gives an optimal fit to the observed \chandra\ spectra (Fig.~\ref{f:spec_sim} lower panel). The "normalized best fit" model is the same as the "best fit" mode in terms of the spectral fit; but the listed $\dot{M}$ values are normalized according to Eq.~\ref{e:norm} from $\delta D=0.014$~pc (used in the "best fit" model simulation) to $\delta D=0.05$~pc, which gives a spatially scaled radial intensity profile that reasonably well matches the \chandra\ data (Fig.~\ref{f:image}B).
\end{table*}

Fig.~\ref{f:spec_sim} illustrates the results of the simulated spectral fits to the observed data. The initial "best guessed" model produces a too flat spectrum (due to the still relatively high wind speed assumed) to fit the observed data well, evident at energies $\gtrsim 4.5$~keV. Although the "best fit" model fits the data reasonably well (with a reduced $\chi^2 = 1.32$), it indicates that the E2 wind speed needs to be cut further in half, compared to the "best guessed" model (Table~\ref{t:para}). Such a low speed value of the WC stellar wind seems to be a bit extreme, in reference to its canonical value. Furthermore, we have here used the X-ray absorbing ${\rm N_H}$ as inferred from the best fit of the \textit{VVAPEC} model to the \chandra\ spectra. As discussed in \S~\ref{ss:res_spec}, the real ${\rm N_H}$ could be 30\% higher. With this higher ${\rm N_H}$, the model spectrum would need to be even softer (i.e., demanding even a lower wind speed) to match the observed ones.

\begin{figure*}
\centering
\includegraphics[width=0.95\linewidth,angle=0]{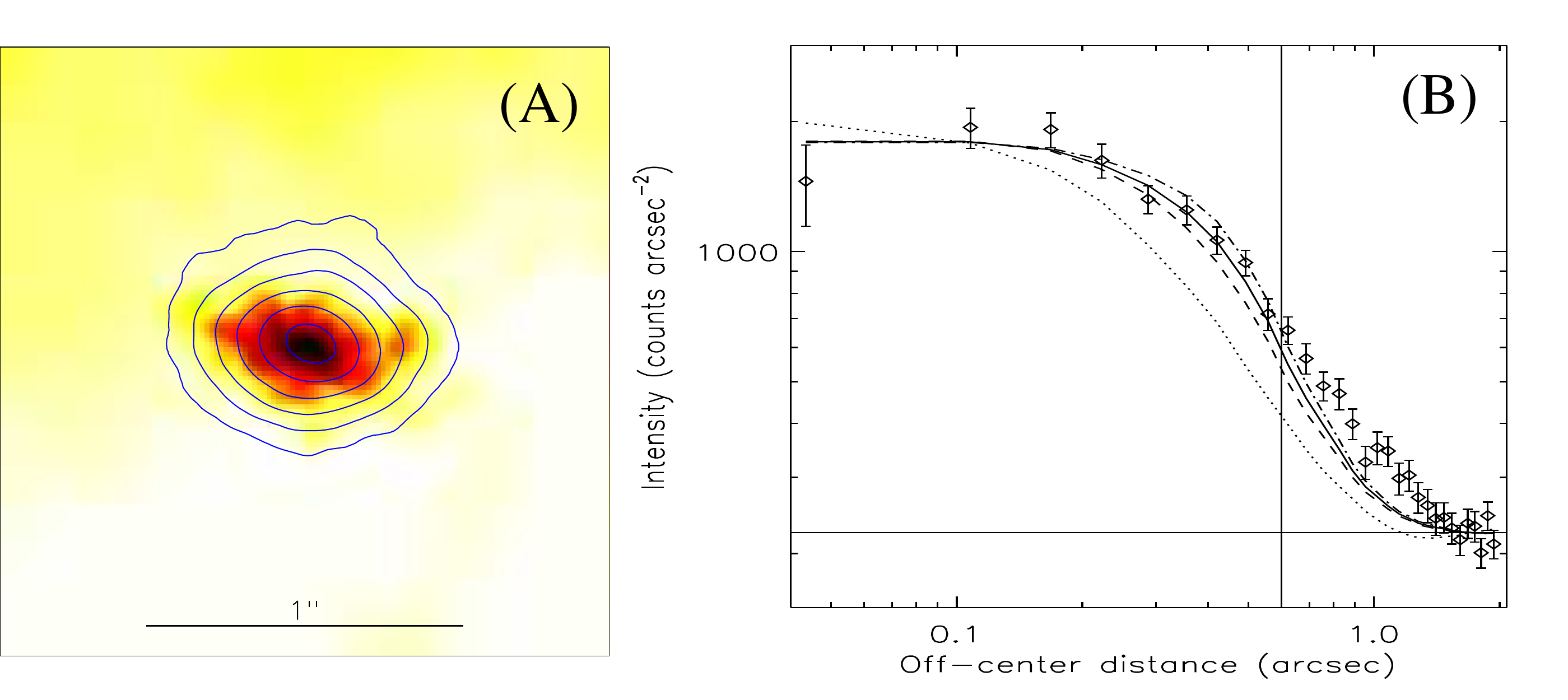}
\caption{(A) Close-up image of the 1-9~keV intensity of \xs, constructed from the ``best fit" model simulations  assuming $\delta D = 1\farcs3$, shown in a logarithmic scale. The overlaid illustrative intensity contours are obtained from the same image after being convolved with the PSF of the \chandra\ observation (see the text). (B) Comparison of the radial model intensity profiles of the PSF-convolved image (solid) with the \chandra\ profile in the 1-9~keV band (data points), as presented in Fig.~\ref{f:f2}A. The vertical line at $r=0\farcs6$ indicates the projected radius or the off-center distance within which the model profile matches the \chandra\ profile well. The horizontal line marks the background level estimated with the \chandra\ data in the 1\farcs5 and 2\arcsec\ range. We have added this level to the model profile after both subtracting its background estimated in the same range and scaling the mean flux of the first three data to the corresponding \chandra\ value. For comparison, similarly constructed model profiles in the 1-4~keV (dashed) and 4-9~keV (dash-dotted) bands, as well as in the 1-9~keV band but assuming $\delta = 0\farcs3$ (dotted) , are also shown.}
\label{f:image}
\end{figure*}

In addition to the spectra, the X-ray synthesis also produces images. Fig.~\ref{f:image}A shows the 1-9~keV emission centered on IRS~13E for the "best fit" model before and after PSF folding. The pre-PSF image is clearly aspherical; the minor axis of the asymmetry is along the line of the two WR stars, so the major axis with a position angle $\approx 80^\circ$ east to the north traces the dominant shocked gas and hence the X-ray emission, closer to E2 than E4, as determined by the ram-pressure balance between the stellar winds. After folding the X-ray models through a \chandra\ PSF, the aspherical nature of the emission is naturally reduced, depending on the assumed $\delta D$. 

The morphology as indicated by the contours shown in Fig.~\ref{f:image}A for our chosen $\delta D \sim 1\farcs3$ (compared to $\delta D=0.014$~pc used in our simulation) resembles that of the emission from the inner region of \xs\ in the \chandra\ image (i.e., the region enclosed by the 3rd contour in Fig.~\ref{f:f1}A). This similarity is also seen in the radial intensity profile comparison within $r \lesssim 0\farcs6$, where a good match to the \chandra\ profile is found for the synthesized one when it is scaled up to $\delta D \sim 1\farcs3$, which is a factor of $\sim 4$ larger than the projected separation between E2 and E4 (Fig.~\ref{f:image}B). Outside this radius, the model systematically under-predicts the intensity. The predicted profile in the 4-9~keV band is only slightly broader than than in the 1-4~keV band. The model profiles are all substantially steeper than the observed one. The intensity excess above the best-matched model here is about 28\% of the total observed flux in the 0\farcs6-1\farcs5 range, but accounts of only about 10\% of the flux within the 1\farcs5 radius. The CW seems to dominate the overall observed emission and the required mass-loss rates of the two WR stars are quite high (Table~\ref{t:WindParam}). This result seems to be quite robust, insensitive to the stellar wind parameter choice. 

With the change of $\delta D$ from 0\farcs34 (used in the simulation) to 1\farcs3, which better matches the observed radial intensity profile, we infer from Eq.~\ref{e:norm} the corresponding $\dot{M}$ values, assuming the same X-ray luminosity (or the spectral normalization) that gives the "best fit" to the \chandra\ spectra. These inferred $\dot{M}$ values are listed under the "normalized best fit" model in Table~\ref{t:para}.

However, the above $\delta D$ scaling represents an upper limit to the extent that the CW X-ray emission extends. $\delta D$ could be significantly smaller than $1\farcs3$. If this is the case, then even a larger fraction of the observed extended emission needs to be explained with a contribution in addition to the CW (see \S~\ref{s:dis} for further discussion).

There are a few other more subtle observed characteristics that are not well accounted for by the simple CW scenario. These features include the apparent offset of the X-ray centroid to the west of the two WR stars and the fan-out morphology of the X-ray emission toward the SE (\S~\ref{ss:res_spat}), as well as the softness of the observed X-ray spectra, compared to the simulated ones. We hypothesize that these characteristics, as well as the observed excess emission in the 0\farcs6-1\farcs5 range, are due to an additional X-ray contribution from the interaction of the \xs\ cluster wind with its ambient medium. We explore this idea below, together with other lines of observational evidence for the interaction.

\section{\alma\ observations of \xs}\label{s:alma}

\subsection{Description of \alma\ observations and data processing}\label{ss:alma-des}

\alma\ observations allow us to  sensitively probe cool gas in the GC, as has been demonstrated in several publications~\citep[e.g.,][]{moser17,tsuboi16,tsuboi17a,tsuboi17,tsuboi19,murchikova19}. Our study here is meant to complement published results and to provide new insights into the spatial and kinematic properties of the gas. While our analysis of relevant observations will be detailed in a separate paper on broader issues of the GC~(Li \etal\ 2019 in preparation), the present study primarily uses the \alma\ band-6 observation taken in May and July, 2017 for the project 2016.1.0087.S, which is focused on the accretion flow toward Sgr A* (PI:  Elena Murchikova). When we started this part of the study, this observation gave us the highest spatial resolution view of the H30$\alpha$ hydrogen recombination line emission of \xs \footnote{Very recently, however, a higher resolution observation has become available (\citealt{tsuboi19};  see also our discussion in \S~\ref{s:exam}).}. No existing publication has used this observation --  \citet{murchikova19} actually used earlier observations.
Our focus here is of course on \xs.

We use the Common Astronomy Software Applications package (CASA) version 5.1.1 for standard data processing. The achieved synthesized beam of the observation is 
$0\farcs22\times 0\farcs13$ (FWHM). We apply the primary beam correction for all the \alma\ images presented in the present work. The final H30$\alpha$ data cube has a spectral resolution of $22~\kms$ and a channel noise of $\sigma_{\rm ch}\approx0.08 {\rm~mJy~beam^{-1}}$, estimated with line-free channels and regions. The moment zero intensity of the H30$\alpha$ line emission, integrated from velocity from $-400~\kms$ to $400 \kms$, has a noise of $\sigma_{\rm mom.0}\approx0.04 {\rm~Jy~beam^{-1}km~s^{-1}}$, while the continuum emission has $\sigma_{\rm cont}\approx 0.1 {\rm~mJy~beam^{-1}}$.

\subsection{\alma\ results}\label{ss:alma-results}

\begin{figure*} 
\centering
\includegraphics[width=1\linewidth,angle=0]{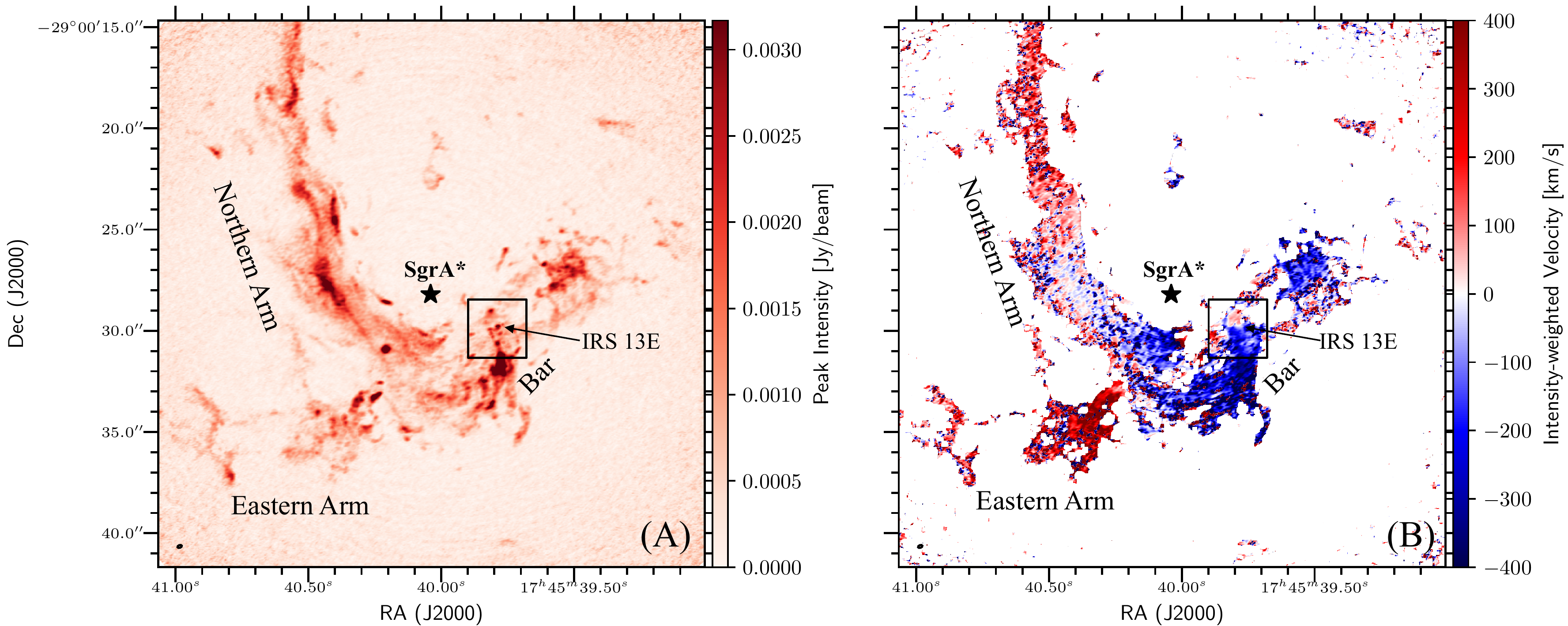}
\caption{
 An \alma\ overview of the GC field: (A) the H30$\alpha$ line peak emission map;  (B) the intensity-weighted velocity map. 
Major ionized gas streams or arms of the minispiral, together with key objects relevant to the present study, are labeled. The small rectangular box marks the \xs\ close-up region shown in Fig.~\ref{f:IRS13E-closeup}.  
}
\label{f:overview}
\end{figure*}

\begin{figure*} 
\centering
\includegraphics[width=1\linewidth,angle=0]{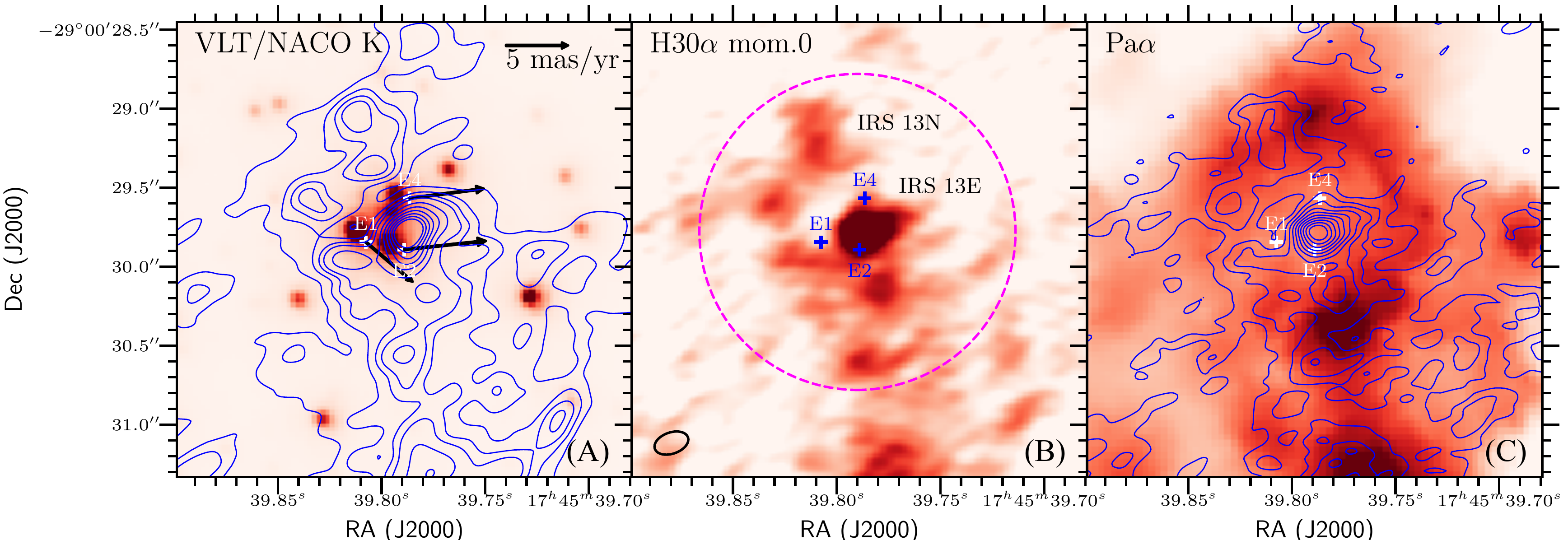}
\caption{
  Close-up views of \xs\ and its immediate surrounding: (A) the \alma\ integrated intensity contours of the 232 GHz continuum overlaid on a {\sl VLT}/NACO K-band image, 
  together with proper motion vectors of IR sources \citep{fritz10};
  (B) \h30a\ line integrated emission intensity map; (C) \h30a\ intensity contours compared with the diffuse \pa\ line emission image~\citep{wang10,dong11}.  
 The beam size is illustrated in the lower-left corner of Panel B.  While the {\sl VLT} IR image in (A) was taken in 2005, 12 years earlier than the \alma\ observation, the positions of E1, E2, and E4 marked with the {\it plus} symbols in all three panels have been corrected for the proper motions ($\sim 0\farcs08$)~\citep{paumard06}.  The magenta circle in (B) outlines a region of 1\as\ radius around E3, from which the \h30a\  flux is extracted in \S~\ref{ss:dis-physical}.}
\label{f:IRS13E-closeup}
\end{figure*}

\begin{figure*} 
\centering
\includegraphics[width=0.95\linewidth,angle=0]{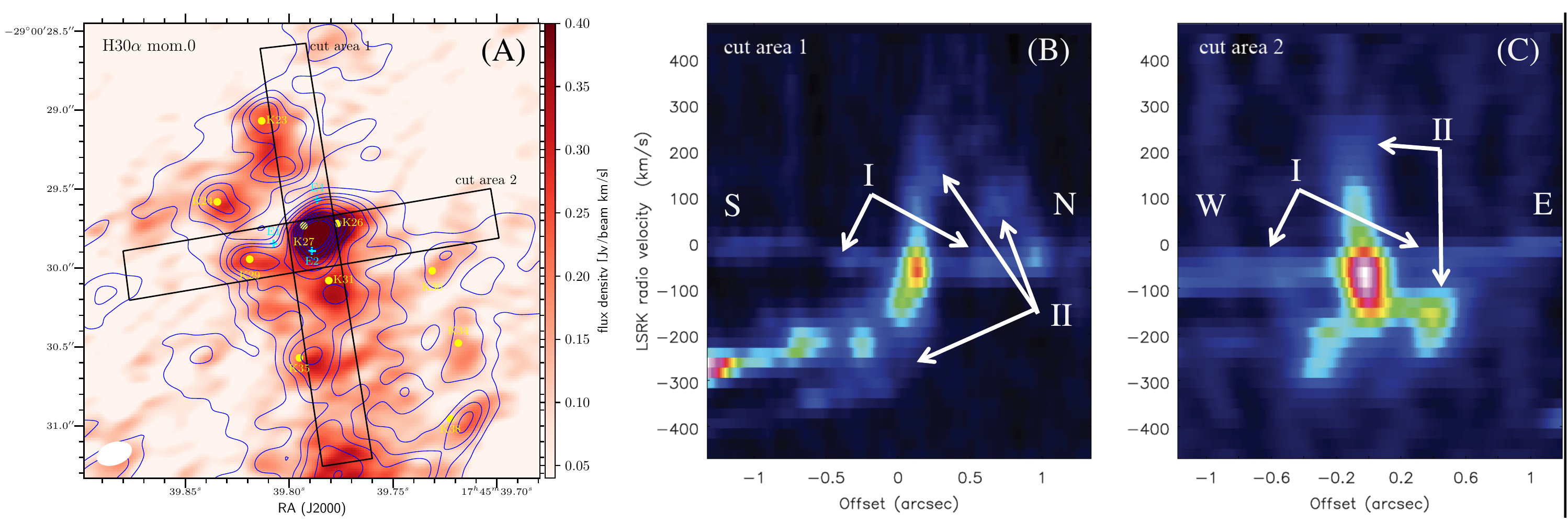}
\caption{
  (A) Close-up of \xs\ and its immediate vicinity in the \alma\ 232 GHz continuum (contours) and \h30a\ line (image) emissions. Compact radio sources taken from~\citet{zhao09}, as well as other key stellar objects in IRS 13E, are also marked with yellow solid dots and plus signs respectively. The two rectangular cut areas (1 and 2) mark the regions through which the position-velocity diagrams, as shown in panels (B) and (C), respectively, are extracted. In these latter panels, the two velocity components (I and II), as well as the directions (S and N, E and W) of the cuts, are marked.}
\label{f:p-v-diag}
\end{figure*}

Fig.~\ref{f:overview} presents an \alma\ overview of the GC field, covering all the major relevant objects and features discussed in the present study, while Figs.~\ref{f:IRS13E-closeup} and \ref{f:p-v-diag}A give close-ups of \xs\ and its immediate vicinity. Fig.~\ref{f:p-v-diag}A  further demonstrates a good correspondence of \alma\ emission peaks with compact radio sources~\citep{zhao09}. These plots show that \xs\ is an outstanding millimeter (mm) wave-emitting source, particularly in terms of the \h30a\  emission. Both the mm continuum and the line emission are centrally peaked with their centroids consistent with each other, located between E2 and E4 (Fig.~\ref{f:IRS13E-closeup}; the two WRs only show weak 232 GHz continuum emission, but no significant H30a emission in the higher resolution image of \citealt{tsuboi19}). This central peak, identified as the near-IR dust emission complex E3~\citep[e.g.,][]{fritz10,tsuboi19}, is rather compact, but still clearly resolved. In addition to the strongly peaked line and continuum emission, interesting surrounding structures are apparent in the plots. We detail the morphological and kinematic properties of these features in the following.

\begin{figure*} 
\centering
\includegraphics[width=1\linewidth,angle=0]{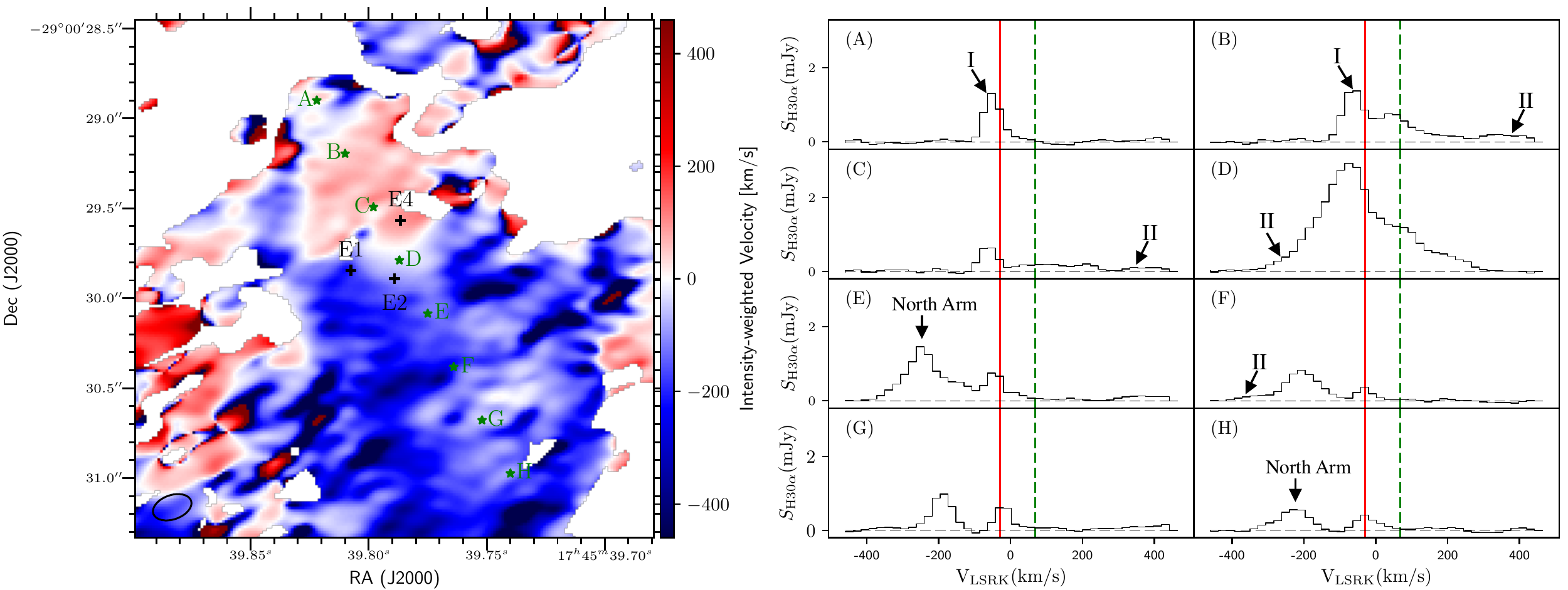}
\caption{
  \h30a intensity-weighted velocity map (left panel) and\ line profiles (right Panels A-H) extracted from the locations marked with the stars along the NE-SW oriented straight line in the left panel.  All the line profiles are obtained within a beam-sized aperture. The solid and dashed vertical lines mark the LoS velocities of the bar and \xs, as listed in Table~\ref{t:IRS13E-para}. The two velocity components (I and II), in addition to the Northern Arm, are marked.
}
 \label{f:v-profs}
\end{figure*}

We confirm the presence of \h30a-emitting gas with a very large LoS velocity spread~\citep{tsuboi17,tsuboi19}.  While Appendix~\ref{a:maps-ch} presents the channel maps of the \h30a\ emission, Figs.~\ref{f:p-v-diag} B and C present the position-velocity diagrams in two particular directions across \xs. Fig.~\ref{f:v-profs} shows the mean LoS velocity structure resolved on scales down to $\sim$ 1600 AU. Gas on the immediate northeastern (NE) side of E3 is primarily redshifted, reaching $\sim +100 \kms$ on average, whereas the blueshifted gas as much as $\sim -200 \kms$ is on the opposite side. Gas with even higher negative velocities appears mostly in regions further south ($\gtrsim 1\arcsec$ away from E3) and is apparently connected to the large-scale components D and C with velocities $\sim -340~\kms$, as identified by~\citet{tsuboi17}. This can also be appreciated in Figs.~\ref{f:p-v-diag}B and C. 

From examining the position-velocity diagrams and the \h30a\ line profiles across the region (e.g., Figs.~\ref{f:p-v-diag} and \ref{f:v-profs}), we can further determine the LoS velocity distribution of the \h30a\ emission. We identify two interesting LoS velocity components. Component I is relatively narrow and centered at $\sim -30 \kms$.  It most probably corresponds to the bar of the minispiral (Table~\ref{t:IRS13E-para}). This narrow bar component seems to be present across the \xs\  region  in Fig.~\ref{f:v-profs}, although confusion with other components is considerable at various locations. The significance of the component in general decreases from the NE to SW. At the location of  \xs\ (see Panel D), the bar component is greatly enhanced in both intensity and LoS velocity spread.

Component II is very broad (with a total spread of $\sim 800 \kms$) and is centered roughly at $\sim +44 \kms$ -- the systemic velocity of \xs\ (Table~\ref{t:IRS13E-para}). This component is seen within about 1\arcsec\ of \xs\ and is in general quite faint ($S_{H30\alpha} \lesssim 0.2$~mJy; Figs.~\ref{f:v-profs} and \ref{f:p-v-diag}), except for at the center of the group. On the NE side of the group, the component is seen essentially only on the redshift side, up to $\sim +420 \kms$ (Panels B-C). But starting at \xs\ (Panel D) and toward the SW, extended blueshifted emission (probably reaching $\sim -320 \kms$) appears, although the velocity spread is very much confused with a distinctly narrow feature, which peaks at $\sim -220 \kms$ (Panels E and F) and represents the continuation of a large-scale coherent structure toward the south, as part of the North Arm (Fig.~\ref{f:overview}B). We do not analyze this feature further in this work.  Nevertheless, Component II is still visible at velocities $\sim -300\kms$ to $-400\kms$, especially in Panel F  (see also Fig.~\ref{f:p-v-diag}B).  Probably because  the \h30a-emitting gas is quite lumpy and/or the sensitivity of the data is too limited (see \S~\ref{ss:dis_struct_large} for discussion),  the component is not apparent in the 2-D channel maps (Fig. ~\ref{f:maps-ch}). Thus its global structure  remains uncertain.

Components I and II probably reflect the influence of \xs\ on the ambient gas in two different regions. Component I is very local to the central region of \xs, between E2 and E4, whereas Components II is over a relatively larger scale up to $\sim 1''$ radius. 
\section{Critical examination of existing results and interpretations}\label{s:exam}

 \citet{tsuboi19} have recently reported a very high-resolution \alma\ observation of \xs\ (beam size $0\farcs037 \times 0\farcs024$). This observation allows them to resolve the central thermal emission peak into multiple features, including extended ones in both the continuum and \h30a\ emissions. The reported work is focused on the brightest extended feature E3, which is characterized with a 2-D Gaussian of  sizes $=0\farcs093 \times 0\farcs061$.  They estimate the density and thermal pressure of the region as $n_e = 6\times 10^5 {\rm~cm^{-3}}$ and $P/k \sim 8 \times 10^9 {\rm~K~cm^{-3}}$. Interestingly, this thermal warm gas pressure is comparable to that inferred for hot gas in \xs\ (\S~\ref{ss:res_spec}), indicating that two gas phases are in rough pressure balance. 
 
 The \h30a\ emission  shows a large velocity gradient, ranging from $\sim -180 \kms$ (at the SE) to $\sim 150 \kms$ (NW). This velocity gradient is nearly opposite to that on larger scales (\S~\ref{ss:alma-results}). \citet{tsuboi19} interpret E3 as a ring-like structure rotating around a dark object and having an inclination angle of $i \sim 50^{\circ}$. The inferred mass of this object is $2.4 \times 10^{4} M_{\sun}$, which could then be the putative IMBH. The structure, however, has to be highly asymmetric or lopsided in its brightness distribution  (see Fig.~4a in~\citealt{tsuboi19}). The NE part of the structure is substantially brighter than the SW. 

 The position and structure of E3 vary with time. A comparison of the \alma\ data with a {\sl VLA} radio image, taken in 1990s and with resolution of $0\farcs09 \times 0\farcs06$, shows an apparent change in the orientation of the emission morphology of E3, which is called HII-7E in~\citet{zhao98}. The radio image shows a NW-SE orientation~\citep[see Fig.~4 in ][]{zhao98}, which is nearly perpendicular to that seen in the \alma\ observation~\citep{tsuboi19}. This systematic orientation change is at least qualitatively consistent with the differential proper motions (predominately in the Declination direction) between the NW and SE parts, labeled as IRS 13W and IRS 13E in~\citet{zhao98}. This work also finds little proper motion of E3 in the RA direction, in sharp contrast to E2 and E4, which show large westward motion (Table~\ref{t:IRS13E-para}; see also Fig.~5 in~\citealt{eckart13}). This difference explains the changing position of E3 relative to E2 and E4, from the west side (in 1990s) to the east side (in 2018;~\citealt{tsuboi19}).
 
 Variability is also seen in the comparison of the \alma\ data with a co-add of near-IR images taken between 2002 and 2007 (see Fig. 1 in~\citealt{fritz10}). In this co-add, E3 can be decomposed into six clumps (E3.0-3.5). The overall morphology of the near-IR E3 emission (e.g., the east-west elongation) is similar to that seen in the \alma\ image (e.g., Fig.~4a; \citealt{tsuboi19}). However, E3 is evidently brighter on its western part (peaking at E3.0) in the co-add, whereas the NE part is enhanced in the \alma\ image (see Fig.~4a; ~\citealt{tsuboi19}). Furthermore, the LoS velocity of E3.0 -- the peak of the E3 complex -- has changed from $\sim -25 \kms$~\citep{fritz10} to $\sim -75 \kms$~\citep{tsuboi19}. Such drastic variations have taken place on a time interval of $\sim 15$~yr between the near-IR and \alma\ observations, for example, which may be allowed by the orbital motion with a period of 50-100~yr in the IMBH interpretation~\citep{tsuboi19}.

However, there might be some difficulties in the IMBH interpretation of E3. First, for a ring around the IMBH, which dominates the gravitational mass of \xs, one would expect an overall velocity spread  symmetry relative to the group. The measured LoS \h30a\ velocity spread of E3, as described above, is centered at $\sim -15 \kms$ (see also Fig.~4 in \citealt{tsuboi19}), apparently inconsistent with those of the three stars in \xs\ (all have positive velocities; Table~\ref{t:IRS13E-para}). 
Because they are located quite some distances away even in projection and on different sides of the putative IMBH, their velocities are not expected to drastically and systemically change over the last 15 years or so. Second, the proper motion pattern of the dusty clumps in E3 shows no indication of rotation~\citep{fritz10}.  
Third, the proposal for the presence of the IMBH in \xs\ is questioned based on measurements of stellar kinematics~\citep{fritz10}. The mass constraint on the IMBH is largely based on upper limits on the expected reflex motion of Sgr A*, as well as the acceleration of E1 and the velocity differences between E1, E2, and E4.
The constraint is greatly weakened if E1 is not bound to \xs. In fact, the velocity vector of E1 is quite different from those of E2 and E4 with an offset of $\sim 174 {\rm~km~s^{-1}}$ (Table~\ref{t:IRS13E-para}). On the other hand, E2, and E4  are most likely bound. The probability for a chance coincidence of these two rare extreme objects to be so close in the phase space is extremely small, especially if they are at the distance of $\sim 10\arcsec$ off Sgr A* (see below and \S~\ref{s:dis}). So we conclude that E2 and E4 are most likely bound, whereas E1 is not a member of the \xs\ group.

Finally, the gravitational bounding between E2 and E4 requires that their separation $\delta D < 2 r_J$, where 
\begin{equation}
    r_J \sim D[m_g/(2M_{Sgr A*})]^{1/3}
\label{e:rj}
\end{equation} 
is the Jacobi radius of \xs\ of a mass $m_g$ against the gravitational tidal force of Sgr A*. 
For the Sgr A* mass $M_{Sgr A*} =4 \times 10^6 ~M_{\sun}$, the orbital distance of the group from Sgr A* $D > 7\arcsec\ (\delta D/1\farcs3)(m_g/10^4 ~M_{\sun})^{-1/3}$. This distance constraint  is consistent with the hypothesis that \xs\ is embedded within the mini-spiral bar, which is distributed within an orbital distance range of 7\arcsec\ -20\arcsec~\citep[][ see also \S~\ref{s:dis}]{liszt03}. The constraint may also be compared with the group's distance of $\sim 7\arcsec\ (V_{E2+4}/249 {\rm~km~s^{-1}})^{-2}$ (Table~\ref{t:IRS13E-para}) for a circular motion; a larger distance is allowed for an eccentric orbit. These arguments seem to favor the presence of an IMBH of mass $\sim 10^4 ~M_{\sun}$. However, such an IMBH is not necessary. With the group's total stellar mass up to $\sim 2000 ~M_{\sun}$~\citep{fritz10}, for example, IRS 13E needs only be $D \gtrsim 10\arcsec(\delta D/1\farcs3)$ away from Sgr A* to avoid tidal disruption. $\delta D$ could also be considerably smaller than our assumed 1\farcs3 here~(see \S~\ref{s:dis}). 

In summary, the presence of the IMBH in \xs\ helps to explain its gravitational bounding but has difficulties with the extreme gas kinematics.  In the next section, we propose an alternative scenario for the gas kinematics.

\section{Collision of \xs\ with ambient dense gas}\label{s:dis}

Energetic stellar winds in \xs\  not only collide with each other, but also unavoidably interact with its surrounding. The impact of this interaction depends on the density of the ambient medium, as well as the relative velocity between the group and the medium. We start by considering their kinematics and will then examine how the collision may explain the thermal emission features with unusually large velocity spreads, as well as various other outstanding phenomena in and around the group. 

\subsection{Kinematics of \xs\ and its surrounding medium}\label{ss:dis-intr}

Table~\ref{t:IRS13E-para} presents a summary of key velocity measurements relevant to the collision of the stellar winds from \xs\  with  the surrounding gas. We first estimate the effective mean velocity of \xs\ ($v_{E2+4}$) by averaging those measured for E2 and E4, which dominate the stellar wind energetics in the group. The result is included in Table~\ref{t:IRS13E-para}. 

Can $v_{E2+4}$ be considered as a representative velocity of \xs?  The answer depends on whether or not the internal velocity dispersion in the group, $\delta V \sim (2G m_g/r_e)^{1/2}$, is substantially smaller than the orbital velocity around Sgr A*, $V_o \sim (2G M_{Sgr A*}/D)^{1/2}$. Using Eq.~\ref{e:rj} and assuming that the effective radius of the group ($r_e$) is comparable to $r_J$, we can approximately express the ratio as
\begin{equation}
    \delta V/V_o \sim 2^{1/6}(m_g/M_{Sgr A*})^{1/3}=0.15(m_g/10^4~M_{\sun})^{1/3}.
\end{equation} 
Therefore, we may reasonably use $v_{E2+4}$ as the systematic velocity of \xs.

We next consider the velocity of the minispiral bar. We have shown in \S~\ref{s:alma} that the \h30a\ emission of E3 peaks around the LoS velocity of the bar, which we assume is due to the collision. To figure out what direction the collision is in,  we need to know the relative transverse motion  between \xs\ and the bar.  To do so, we use proper motion measurements of various thermal gas blobs in and around \xs~\citep[e.g.,][]{zhao09}.  These measurements, made with high-resolution {\sl VLA} observations, show that proper motion vectors point in many different directions, which may be largely due to the dispersing of the blobs by the stellar winds. We average the vectors to estimate the undisturbed motion of the bar. We find that the averaged vector is primarily toward the west with a speed of $v_{\rm RA} \sim -150 \kms$. In the Declination direction, however, the averaged vector is short,  varies from one work to another and can be positive or negative~\citep[e.g.,][]{zhao99,zhao09}. Therefore,  we approximate $v_{\rm Decl.} \sim 0 \kms$ for the bar in Table~\ref{t:IRS13E-para}. 

Recent studies suggest that the bar is a distinct gaseous structure orbiting around Sgr A*. \citet{liszt03} shows that the bar can be reasonably well represented by a kinematic model of a series of thin co-planar rings rotating at the local circular velocity around Sgr A*. This set of rings is assumed to extend from 0.3 to 0.5 pc in radius and to be  nearly edge-on  and  perpendicular to the Galactic plane. \citet{tsuboi17a} favor this near-polar ring interpretation, but suggest that the orbits of the rings may be highly eccentric. In both models, the bar at the location of \xs\  is largely moving from the east to the west and just changes the LoS component from redshift to blueshift. Within their large uncertainties, the model predictions are consistent with the velocities that we have adopted for the bar at \xs. 

\subsection{Scenario for the collision between \xs\ and its ambient gas}\label{ss:dis-model}

\begin{figure}
\centering
\includegraphics[width=0.95\linewidth,angle=0]{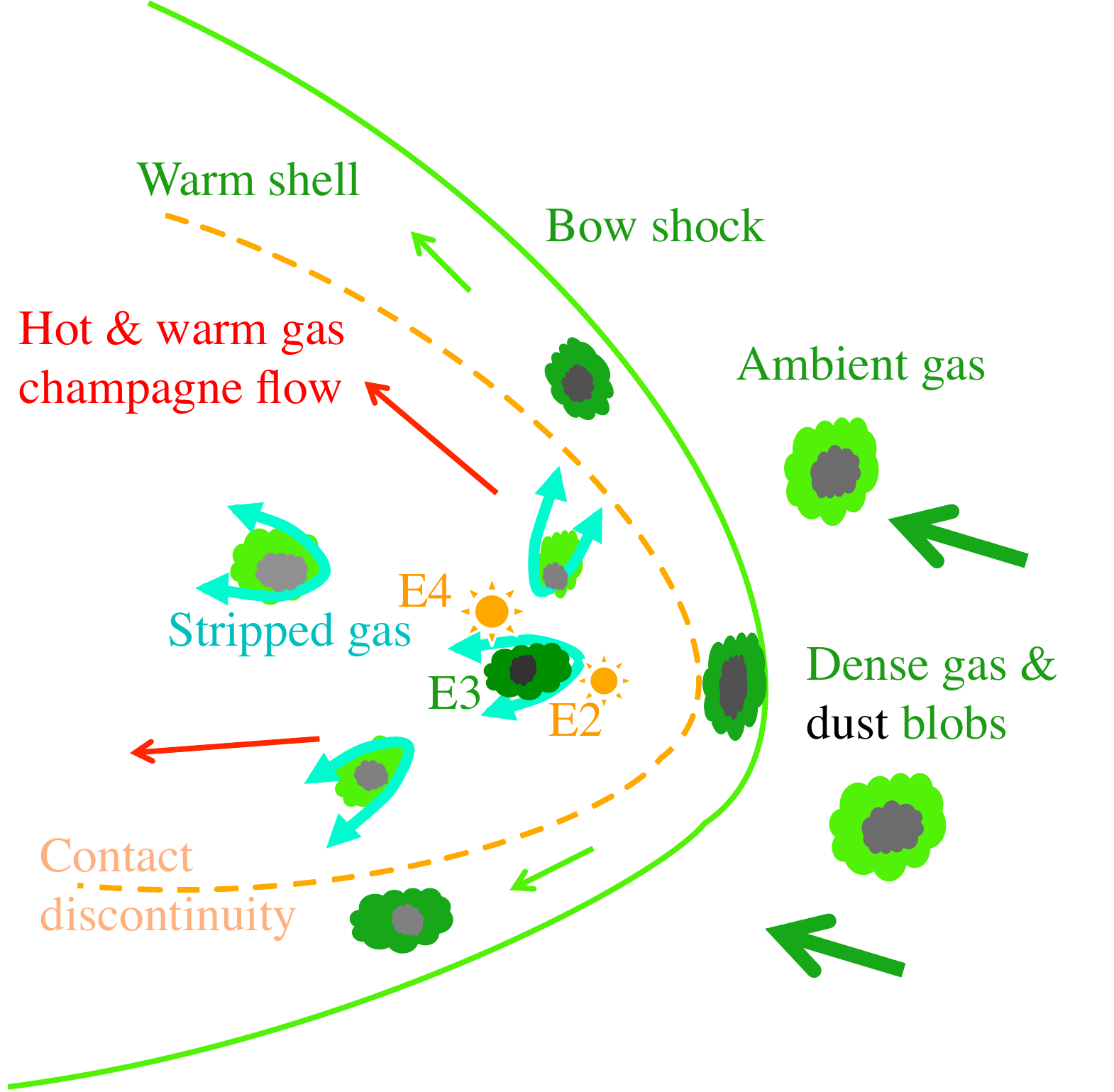}
\caption{
Illustration of the collision scenario between \xs\ and the ambient medium through the plane of E2, E4, and the observer (roughly to the far left). Arrows mark the directions of gas motion relative to the mean velocity of E2 and E4. The illustrated objects are not meant to be properly scaled. }
\label{f:ill}
\end{figure}

With the estimates of the relative motion between \xs\ and the bar, we can now check what kind of collision they may be experiencing and what it could produce. With a  speed of $\sim 1.4 \times 10^2 \kms$ (Table~\ref{t:IRS13E-para}), this collision should be supersonic. We thus expect that a bow shock forms around \xs, as illustrated in Fig.~\ref{f:ill}.  The morphology of this bow shock depends on the exact 3-D locations of the stars, especially those  (E2 and E4) with strong stellar winds, and their  energetics, as well as the direction of the collision, all of which are quite uncertain. Nevertheless, with our rough estimates of the velocity vector of the relative motion ($v_{\rm RA}, v_{\rm Decl.}, v_z =  -88, -39, +98 \kms$; Table~\ref{t:IRS13E-para}),  the collision seems mostly along the LoS direction and toward the west (slightly to the south). 
The high-speed collision, together with the strong ionizing radiation and stellar winds of the group, should also produce an energetic champagne flow in the opposite direction of the relative motion. 

We have examined existing works to explore the characteristics of the bow shock and champagne flow. There are various models that incorporate the effect of ionizing radiation and stellar winds. Such models have been explored in 2-D simulations~\citep{arthur06}, which show limb-brightened morphology of warm gas,  instabilities in the swept-up shell with density fluctuations, and champagne flows with complicated kinematics. However, the collision between \xs\ and the bar is far more extreme than what are assumed in the existing modeling, in terms of the relative velocity and gas density, as well as the structures of \xs\ and the ambient medium. So we expect that the reality is far more complicated than the models set up for the simulations. In particular, the dense medium is likely to be highly structured (see similar case in~\citealt{veena17}). Dense gas blobs, perhaps present in the bar,  need time to be dispersed even in the harsh environment of the bow shock around \xs. In the process, they are compressed by the high pressure in the region (Fig.~\ref{f:ill}). The resultant high density dusty cores may be detected as clumps seen in the near-IR images, while the ionized gas as thermal emission features seen in ratio, mm, and Br-$\gamma$ line emission. On average, incoming blobs are first slowed down in the upper stream (in the region ahead of E2 and E4), against the ram-pressure, and are then pushed by ram pressure to the downstream, forming a champagne flow in the opposite direction of the bow shock. This flow could reach a velocity comparable to the sound speed of the shocked stellar wind material [$5.1 \times 10^2 {\rm~km~s^{-1}} ({\rm~kT/keV})^{1/2}$; ~\citealt{vanburen88,wang93}]. The LoS component of this velocity should then be $\sim 365\kms$.
Given this collision scenario of \xs\ with the bar (Fig.~\ref{f:ill}),  we check in the following how the observed key characteristics in the region may be explained.

\subsection{Emission structure within \xs}\label{ss:dis-peak}

The presence of dusty clumps in \xs\ has been known for quite some time~\citep[e.g.,][]{fritz10}. They are among the brightest and hottest non-stellar objects in the central parsec of the Galaxy and are clearly heated  by the stars in the group. One possible origin of these clumps may be massive ejections from a star (e.g., E4), which could have left the luminous blue variable phase a short time ago~\citep{coker02}. However, this scenario has multiple difficulties. 1) Such clumps should probably be distributed with some symmetry and have outward proper motions, the speed of which increases with the distance from the star because the entrainment by the fast stellar wind. The observed proper motions of the clumps show no such trends. 2) The time scale for the clumps to be advected out of \xs\ by the wind, if the entrainment is efficient, is only $\sim 10 {\rm~yr~v_{w,3}^{-1}}$, which may be far too short statistically to observe the clumps so near their star. 3) The velocity offset of E3.0 from E2 and E4 ($\sim 184 \kms$) is substantially greater than that from the bar ($\sim 88 \kms$). So E3.0 is more likely to originate from the bar than either of the two stars~\citep[see also ][]{fritz10}. 

In \S~\ref{ss:alma-results}, we have already mentioned the apparent connection between the central thermal emission peak (as seen in the \alma\ observations) and the dusty clumps (observed in near-IR; \citealt{fritz10}). They show overall positional and morphological similarities, but also exhibit  structural and kinematic variations. We speculate that the central thermal gas blobs and dusty clumps are physically associated. The dusty clumps are likely embedded within the warm gas blobs, which are heated by the stars. In our group wind collision scenario (\S~\ref{ss:dis-model}; Fig.~\ref{f:ill}), the blobs represent the compressed dense substructures originating in the dense gas bar, similar to those seen in nearby HII regions ~\citep[e.g.,][]{veena17}. The expected velocity evolution of the blobs naturally explains the observed variations in their kinematics: the systematic acceleration of the motion from west to east~\citep{fritz10} and the increasing blueshift of E3.0 with time (see \S~\ref{s:exam}). Individual blobs are expected to suffer evaporation by ionizing radiation and stripping by fast-moving stellar wind material, which could account for the observed large LoS velocity spread. Both the pressure and the ionizing radiation intensity should be the highest in the region between E2 and E4, where blobs are expected to radiate brightly. 
Moreover, the rough thermal pressure balance (\S~\ref{s:exam}) is consistent with the interpretation of the mm and X-ray thermal emissions as arising from distributed and intermixed warm and hot gases in our scenario.

We have also considered the possibility for the formation of warm dusty gas directly from the cooling of shocked stellar wind material \citep{calderon16, calderon19}. However, the separation between E2 and E4 is too large to reach high enough densities for the shocked wind material to cool down significantly, let alone to form dust.

\subsection{Emission structures around \xs}\label{ss:dis_struct_large}

We first consider the observed X-ray emission from \xs. In \S~\ref{s:res}, we have shown that the emission is predominantly thermal and most likely arises from the CW of the massive stars (chiefly E2 and E4)  in the group. The characteristic temperature of 1.34 keV is well consistent with the expected speeds of the winds, and the pressure of the X-ray-emitting region is also consistent with that estimated in the warm dusty gas feature E3. However, the CW between the two stars alone has difficulties to explain the overall extent, fan-out morphology and softness of the X-ray emission, indicating that additional physical processes need to be accounted (\S~\ref{s:res}).

Most naturally, we expect that the interaction of the group wind of \xs\ with the surrounding gas plays a role (Fig.~\ref{f:ill}). We check how  the interaction may be reflected in the X-ray properties of the group. First, the interaction should result in a partial confinement of the wind by the bow shock. Because of this confinement, the wind is expected to be reverse-shocked on the scale $r_s \sim (0.05{\rm~pc}) \dot{M}_{w,-4}^{1/2} v_{w,8}^{1/2} n_{a,5}^{-1/2} v_{r.2}^{-1}$ 
~\citep{vanburen88,draine11}. The contact discontinuity of the reverse-shocked stellar wind material and bow shocked ambient gas should have a scale $r_c$ that is a factor of $\sim 1.5$ greater than $r_s$~\citep{vanburen88}.
We can reproduce the size of the shell ($r \sim 1\arcsec$) by approximating the combined stellar wind power in \xs\ (Table~\ref{t:WindParam}) as a single source and adopting $v_r$ value in Table~\ref{t:IRS13E-para} and $n_{a,5} \sim 1$, consistent with the value estimated for the bar~\citep{liszt03,tsuboi17a}.
In reality, the structure of the contact discontinuity should certainly be considerably more complicated and should be elongated in the projected direction connecting E2 and E4. Second, the pressure gradient from the bow shock to the champagne flow should also result in a fan-out morphology of the X-ray emission, just as observed. Similarly, the emission from the CW between E2 and E4 should peak on the side facing the bow shock, which may explain the slightly westward offset ($\sim 0\farcs15$; \S~\ref{ss:res_spat}) of the X-ray emission peak. Third, the mass-loading of warm ionized gas into the hot gas can occur around individual dense blobs and at the contact discontinuity of the bow shock. The mass-loading naturally reduces the temperature of the hot gas, which should produce a softer spectrum than what is expected from the simple CW, as observed. If most of the extended X-ray emission is due to the cluster wind interaction with the bar, then the required increase of $\delta D$ can be reduced (see \S~\ref{s:simul}). Correspondingly, $\dot{M}_{w}$ can also be lower than those quoted in Table~\ref{t:WindParam} (the "best fitted" and $\delta D=1\farcs3$ case). 
In short, the spatial and spectral properties of \xs\ can be better explained by accounting for the interaction of the stellar winds with the ambient medium, in addition to the CW between E2 and E4.
           
We next consider the thermal emission from the ambient medium of \xs.
In the hydrogen \pa\ line emission map of the GC~\citep{wang10}, we notice a shell-like structure around \xs\ (Fig.~\ref{f:IRS13E-closeup}C), in addition to its own emission which has been excised~\citep{dong12}.  This structure has an overall size of $\sim 2''$, comparable to that of the diffuse X-ray emission, is centered roughly at \xs, and is  elongated in the N-S direction, 
which is consistent with the orientation of E2 and E4. Therefore, most likely the structure represents the limb-brightened density enhancement of photo-ionized warm gas in the down-stream of the bow shock (Fig.~\ref{f:ill}). Furthermore, the western part of the structure is generally brighter and better defined than the eastern one, consistent with the bow shock morphology expected from the westward motion of \xs\ relative to the ambient medium in the bar.

The \pa\ emission is relatively brighter than the average at the northern and southern ends. The shell-like structure is more well-defined on the western side than on the eastern side. However, the opposite is apparent in the \h30a\ image.  This difference may be explained by the differential attenuation of the \pa\ emission: higher on the eastern side than on the western one, while the \h30a\ suffers no such effect. The \h30a\ image shows no clear detection of the western shell-like structure,  although patchy emission with large velocity offset and broadening is present (Component II discussed in \S~\ref{ss:alma-results}). This apparent overall morphological difference may be due to its lower detection sensitivity than the \pa\ one. The \pa\ intensity of the structure, as well as the apparent compact feature about 0\farcs7 further west, is $\lesssim 2\sigma$ above the background in the \h30a\ image. The noise may also partly explain the lumpy appearance of the image, compared with the \pa\ image  (see also \S~\ref{ss:alma-results}). Therefore, we tentatively suggest the presence of a shell-like structure, which is probably intrinsically brighter on the eastern side than on the western side.

In short, the CW and the ram-pressure confinement of the group wind by the ambient gas can naturally explain the structural, kinematic, and variability properties of dusty gas, as well as the X-ray emission, in and around \xs.

\subsection{Physical properties of warm ionized gas in \xs}
\label{ss:dis-physical}
With the \alma\ results, we now estimate such physical parameters as the temperature, density, and volume filling factor of warm ionized gas in the \xs\ region. Assuming that that this gas is under the local thermodynamic equilibrium and its 232 GHz continuum and \h30a line emissions are optically thin, the electron temperature $T_{\rm e}$ can be expressed as
\begin{equation}
T_{\rm e}({\rm K}) = \left[ \frac{6985}{\alpha (\nu,T_{\rm e})} \left( \frac{\nu}{\rm GHz}\right)^{1.1} \frac{1}{1+N({\rm He})/N({\rm H})}\times \left( \frac{S_{\rm C}}{S\Delta V_{\rm H30\alpha}} \right)  \right]^{0.87},
\end{equation}
where the elemental number density ratio of $N(\rm He)/N(\rm H) \sim 0.08$ and $\alpha(\nu,T_{\rm e})\sim 0.85$ at $\nu=232$ GHz and $T_{\rm e} \sim 10^4$ K \citep{mezger67}. For E3, the total flux densities of the continuum and \h30a line emissions are $S_{\rm C}=28.5\pm 1.8~{\rm mJy}$ and $S\Delta V_{\rm H30\alpha}=2.37\pm 0.12~{\rm Jy~km~s^{-1}}$. We  thus have $T_{\rm e}=9260\pm 650$ K, which is consistent with previous estimates, ranging from 6000 K to 12000 K~\citep{moser17,zhao10, tsuboi19}. We can also obtain the volume emission measure (EM)  as~\citep{murchikova19}
\begin{equation}
    {\rm EM} = S\Delta V_{\rm H30\alpha}\frac{4\pi D^2}{\epsilon_{\rm H30\alpha}(T_{\rm e},n_{\rm e})} \frac{\nu}{c} \sim 1 \times 10^{60}\rm~cm^{-3},
\end{equation}
where the \h30a\ emissivity $\epsilon_{\rm H30\alpha}=1.25\times 10^{-31} \rm~erg~s^{-1}$ for $T_{\rm e}=10^4$ K and $n_{\rm e}=10^4\sim 10^7 \rm~cm^{-3}$ \citep{storey95} and the speed of light $c$ are used. Furthermore, we find that both the continuum and \h30a emission intensities of E3 can be characterized by a Gaussian with a deconvolved size of $\sim 0.01$~pc, assuming a spherical symmetry. This together with the above EM value leads to the estimate of $n_{\rm e} \sim 3 \times 10^5\rm~cm^{-3} f_V^{-1/2}$. By further assuming a thermal pressure balance between the warm and hot gases (e.g. $P/k \sim 2 \times 10^{10} {\rm~K~cm^{-3}}$; \S~\ref{ss:res_spec}), we estimate  the volume filling factor $f_V \sim 0.1$ and hence $n_{\rm e} \sim 10^6\rm~cm^{-3} $. If  similar temperature and pressure could be assumed for warm ionized gas on larger scales, its filling factor has to be very small.  We estimate that the total \h30a\  flux within a circular region of 1\as\ radius centered on E3 (Fig.~\ref{f:IRS13E-closeup}B) is $\sim 12 \rm~Jy~km~s^{-1}$, which corresponds to ${\rm EM} = 6\times 10^{60}\rm~cm^{-3}$. The filling factor is then $\sim 8\times 10^{-4}$, while the mass of the warm ionized gas in this region is $\sim m_p EM/n_e\sim 1 \times 10^{-2}~M_\odot$. These estimates indicate that the volume is indeed dominated by hot gas and that warm gas traces dense clumps, probably their surfaces exposed to the ionizing radiation from massive stars in the region.

\subsection{Potential effects of the group wind on larger scales}

 The interaction of the \xs\ group wind with the surrounding medium should further occur on scales greater than the bow shock. The champagne flow, consisting of dusty warm clumps and hot gas, will eventually be terminated by the ram-pressure generated collectively by the outflow of Sgr A*~\citep{wang13}, as well as winds from other stars in the GC cluster~\citep{paumard06}.  A quantitative study, however, requires a proper modeling of the champagne flow, which is not easy to do, because of its multi-phase nature. Nevertheless, we expect that a  shock should terminate and eventually disperse the flow. A plausible intermediate product of this process is IRS 13N -- a group of co-moving dusty objects located only about 0\farcs5 northeast of \xs~\citep{eckart13}.

 Of course, high-speed collisions as experienced by \xs, can also occur within other massive stars or groups. Such collisions may be a viable mechanism for dispersing dense gas in known organized streamers (Fig.~\ref{f:overview}). Gas and dust clumps or complexes have been observed to be scattered around Sgr A*~\citep[e.g.,][]{eckart13,goicoechea18}. Their velocities generally do not match those of the streamers and arms. Such off-stream clumps could well represent remnants of dense gas blobs dispersed by similar encounters with massive stars.
 
\section{Summary}

We have conducted a detailed multi-wavelength study of \xs\ and its vicinity, based primarily on archival \chandra\ and \alma\ observations, as well as dedicated hydrodynamic simulations of the CW scenario. Our main results and conclusions are as follows:

\begin{itemize}
\item We find no significant change in X-ray properties of \xs\ over a period of 15 years. The X-ray spectrum of \xs, displaying distinct emission lines such as S, Si and Fe He-$\alpha$ transitions, can be well fitted with an optically-thin thermal plasma of a characteristic temperature $\sim 1.5 \times 10^7$~K. The X-ray emission, with a 2-10 keV luminosity of $ \sim 1.2 \times 10^{33} {\rm~ergs~s^{-1}}$, is well resolved and has a size of about 2$^{\prime\prime}$. The X-ray emission has a fan-out global morphology; its centroid may be slightly offset ($\sim 0\farcs15$) from the effective center of the group toward the west -- the direction of the group's proper motion.

\item   No single known  massive star (or potentially binary) in the group seems to dominate the X-ray emission. Assuming that a point-like source coincides spatially with the central peak of the emission, we constrain its luminosity as $ \lesssim 6 \times 10^{32} {\rm~ergs~s^{-1}}$ in the 2-10 keV band. A more physical  consideration, based on an analogy to Sgr A*, shows that the luminosity of  the IMBH, if present and undergoing hot accretion, should be orders of magnitude fainter.

\item We detect for the first time the X-ray emission from the eclipsing X-ray binary E60, just 1\farcs3 away from \xs.  The emission also appears to arise from the CW of this binary. We find no indication for any interaction of the wind-wind collision between E60 and \xs, indicating that they are not physically associated. This is consistent with the large LoS velocity difference between the two stellar systems.

\item  The X-ray emission of \xs\ is dominantly due to shock-heated stellar winds from the massive stars, chiefly the two well-separated WR stars (E2 and E4), in the group. This CW scenario is consistent with the X-ray emission's peak position between the WR stars.  A comparison of the X-ray data with the simulations broadly confirms the scenario and provides interesting constraints on the wind parameters, as well as the physical separation between E2 and E4. The separation could be up to $\sim 1\farcs3$ (or 0.05~pc), about four times greater than the projected one, which would account for the bulk of the observed X-ray luminosity with quite nominal wind parameters. Nevertheless, even the best-matched model systematically under-predicts the X-ray intensity in $r \gtrsim  0\farcs6$ regions. The predicted spectral shape is also somewhat harder than the observed, especially if the extinction-inferred high X-ray-absorbing gas column density is considered. 

\item From the consideration of the tidal radius of \xs\ against the gravity of Sgr A*, we conclude that E1 is most likely not bound to the \xs\ group, while E2 and E4 are physically associated. The combination of the gravitational bounding and the above best-matched separation of the two stars further suggests that the orbital distance of \xs\ from Sgr A* is greater than 0.27~pc (equivalent to 7\arcsec) and/or that the mass of the group is primarily non-stellar. If the separation is much smaller than $\sim 1\farcs3$, then the extended X-ray emission needs to be primarily produced by the interplay between the \xs\ group wind and the ambient medium located in the so-called bar of the minispiral around Sgr A*.  
  
\item  The mm emission associated with \xs\ is thermal and is well resolved spatially. The emission within the group is dominated by E3 -- a complex of dusty clumps initially detected in near-IR and radio. Kinematically, this complex exhibits a broad LoS velocity spread up to $\sim 740 \kms$. 
The E3 complex further shows variation in both morphology and LoS speed on a 10 year time scale. 

\item We tentatively identify a shell-like \pa-emitting structure of a dimension $\sim 2''$ and elongated in the north-south direction. This structure can be interpreted as the limb-brightened projection of the bow shock produced by the motion of \xs\ relative to the bar. While the \pa\ emission is enhanced on the western side of the group, the diffuse \h30a\ emission is on the eastern side. 
  
\item  The impact of \xs\ on the environment of Sgr A* is dynamic. With its
  energetic wind, \xs\ can disturb the environment violently. Its strong
  ram-pressure can disperse dense gas in streams or arcs, potentially
  leading to sporadic cool gas inflows that feed Sgr A*~\citep[e.g.,][]{murchikova19}.

\end{itemize}

These new constraints or insights on the presence of an IMBH, the stellar wind properties of the Wolf-Rayet stars, and their physical separation in \xs, as well as its position relative to the minispiral bar and Sgr A*, are important for understanding the environment of Sgr A*. High spatial and timing resolution monitoring of the proper motions and kinematics of the region can be particularly useful to improve the constraints. Further modeling and simulations are needed to better quantify the results, especially the dynamics of the bar and its interplay with \xs, and to further test the scenarios. Similar compact high-velocity features of dense gas observed in the GC may also be explained by such interplay with massive stellar winds -- a phenomenon that may be quite common in galactic nuclear regions, where relative velocity between stellar groups and ambient medium can be very large, and may provide a mechanism to feed SMBHs sporadically with dense gas.

\section*{Acknowledgments}

We thank the referee for constructive comments, which helped to improve the presentation of the paper. QDW is grateful to the financial support and hospitality that he received in the Instituto de Astrof\'i­sica at Pontificia Universidad Cat\'olica de Chile, where part of this work was performed, as well as the Fulbright U.S. Scholar Program for a fellowship. Support for this work was provided by the National Aeronautics and Space Administration through Chandra Award Number GO9-20023X issued by the Chandra X-ray Center, which is operated by the Smithsonian Astrophysical Observatory for and on behalf of the National Aeronautics Space Administration under contract NAS8-03060, and by the ADAP grant NNX17AL67G. JL acknowledges support by the China Scholarship Council (No.201706040153) for the study at University of Massachusetts, Amherst. CMPR acknowledges support from \textsc{FONDECYT} grant 3170870.  CMPR and JC acknowledge support from \textsc{conicyt} project Basal AFB-170002, and from the Max Planck Society through a ``Partner Group'' grant.  Resources supporting this work were provided by the NASA High-End Computing Program through the NASA Advanced Supercomputing Division at Ames Research Center.

This paper makes use of the following ALMA data: ADS/JAO.ALMA\#2016.1.00870.S. ALMA is a partnership of ESO (representing its member states), NSF (USA) and NINS (Japan), together with NRC (Canada), MOST and ASIAA (Taiwan), and KASI (Republic of Korea), in cooperation with the Republic of Chile. The Joint ALMA Observatory is operated by ESO, AUI/NRAO and NAOJ. The National Radio Astronomy Observatory is a facility of the National Science Foundation operated under cooperative agreement by Associated Universities, Inc.
\bibliographystyle{mnras}
\bibliography{GC,sgras}{}

\begin{thebibliography}{}
\makeatletter
\relax
\def\mn@urlcharsother{\let\do\@makeother \do\$\do\&\do\#\do\^\do\_\do\%\do\~}
\def\mn@doi{\begingroup\mn@urlcharsother \@ifnextchar [ {\mn@doi@}
  {\mn@doi@[]}}
\def\mn@doi@[#1]#2{\def\@tempa{#1}\ifx\@tempa\@empty \href
  {http://dx.doi.org/#2} {doi:#2}\else \href {http://dx.doi.org/#2} {#1}\fi
  \endgroup}
\def\mn@eprint#1#2{\mn@eprint@#1:#2::\@nil}
\def\mn@eprint@arXiv#1{\href {http://arxiv.org/abs/#1} {{\tt arXiv:#1}}}
\def\mn@eprint@dblp#1{\href {http://dblp.uni-trier.de/rec/bibtex/#1.xml}
  {dblp:#1}}
\def\mn@eprint@#1:#2:#3:#4\@nil{\def\@tempa {#1}\def\@tempb {#2}\def\@tempc
  {#3}\ifx \@tempc \@empty \let \@tempc \@tempb \let \@tempb \@tempa \fi \ifx
  \@tempb \@empty \def\@tempb {arXiv}\fi \@ifundefined
  {mn@eprint@\@tempb}{\@tempb:\@tempc}{\expandafter \expandafter \csname
  mn@eprint@\@tempb\endcsname \expandafter{\@tempc}}}

\bibitem[\protect\citeauthoryear{{Arnaud}}{{Arnaud}}{1996}]{arnaud96}
{Arnaud} K.~A.,  1996, in {Jacoby} G.~H.,  {Barnes} J.,  eds,  ASP Conference
  Series Vol. 101, ADASS V. p.~17

\bibitem[\protect\citeauthoryear{{Arthur} \& {Hoare}}{{Arthur} \&
  {Hoare}}{2006}]{arthur06}
{Arthur} S.~J.,  {Hoare} M.~G.,  2006, \mn@doi [\apjs] {10.1086/503899}, \href
  {https://ui.adsabs.harvard.edu/abs/2006ApJS..165..283A} {165, 283}

\bibitem[\protect\citeauthoryear{{Baganoff} et~al.,}{{Baganoff}
  et~al.}{2001}]{baganoff01}
{Baganoff} F.~K.,  et~al., 2001, \mn@doi [\nat] {10.1038/35092510}, \href
  {http://adsabs.harvard.edu/abs/2001Natur.413...45B} {413, 45}

\bibitem[\protect\citeauthoryear{{Banerjee} \& {Kroupa}}{{Banerjee} \&
  {Kroupa}}{2011}]{banerjee11}
{Banerjee} S.,  {Kroupa} P.,  2011, \mn@doi [\apjl]
  {10.1088/2041-8205/741/1/L12}, \href
  {http://adsabs.harvard.edu/abs/2011ApJ...741L..12B} {741, L12}

\bibitem[\protect\citeauthoryear{{Bartko} et~al.,}{{Bartko}
  et~al.}{2009}]{bartko09}
{Bartko} H.,  et~al., 2009, \mn@doi [\apj] {10.1088/0004-637X/697/2/1741},
  \href {http://adsabs.harvard.edu/abs/2009ApJ...697.1741B} {697, 1741}

\bibitem[\protect\citeauthoryear{{Calder{\'o}n}, {Ballone}, {Cuadra},
  {Schartmann}, {Burkert}  \& {Gillessen}}{{Calder{\'o}n}
  et~al.}{2016}]{calderon16}
{Calder{\'o}n} D.,  {Ballone} A.,  {Cuadra} J.,  {Schartmann} M.,  {Burkert}
  A.,   {Gillessen} S.,  2016, \mn@doi [\mnras] {10.1093/mnras/stv2644}, \href
  {https://ui.adsabs.harvard.edu/abs/2016MNRAS.455.4388C} {455, 4388}

\bibitem[\protect\citeauthoryear{{Calder{\'o}n}, {Cuadra}, {Schartmann},
  {Burkert}, {Prieto}  \& {Russell}}{{Calder{\'o}n} et~al.}{2019a}]{calderon19}
{Calder{\'o}n} D.,  {Cuadra} J.,  {Schartmann} M.,  {Burkert} A.,  {Prieto} J.,
    {Russell} C. M.~P.,  2019a, arXiv e-prints, \href
  {https://ui.adsabs.harvard.edu/abs/2019arXiv190604181C} {p. arXiv:1906.04181}

\bibitem[\protect\citeauthoryear{{Calder{\'o}n}, {Cuadra}, {Schartmann},
  {Burkert}  \& {Russell}}{{Calder{\'o}n} et~al.}{2019b}]{calderon20}
{Calder{\'o}n} D.,  {Cuadra} J.,  {Schartmann} M.,  {Burkert} A.,   {Russell}
  C. M.~P.,  2019b, arXiv e-prints, \href
  {https://ui.adsabs.harvard.edu/abs/2019arXiv191006976C} {p. arXiv:1910.06976}

\bibitem[\protect\citeauthoryear{{Capelli}, {Warwick}, {Porquet}, {Gillessen}
  \& {Predehl}}{{Capelli} et~al.}{2012}]{capelli12}
{Capelli} R.,  {Warwick} R.~S.,  {Porquet} D.,  {Gillessen} S.,   {Predehl} P.,
   2012, \mn@doi [\aap] {10.1051/0004-6361/201219544}, \href
  {http://adsabs.harvard.edu/abs/2012A%26A...545A..35C} {545, A35}

\bibitem[\protect\citeauthoryear{{Coker} \& {Pittard}}{{Coker} \&
  {Pittard}}{2000}]{coker00}
{Coker} R.~F.,  {Pittard} J.~M.,  2000, \aap, \href
  {https://ui.adsabs.harvard.edu/abs/2000A&A...361L..13C} {361, L13}

\bibitem[\protect\citeauthoryear{{Coker}, {Pittard}  \& {Kastner}}{{Coker}
  et~al.}{2002}]{coker02}
{Coker} R.~F.,  {Pittard} J.~M.,   {Kastner} J.~H.,  2002, \mn@doi [\aap]
  {10.1051/0004-6361:20011745}, \href
  {https://ui.adsabs.harvard.edu/abs/2002A%26A...383..568C} {383, 568}

\bibitem[\protect\citeauthoryear{{Cuadra}, {Nayakshin}  \& {Martins}}{{Cuadra}
  et~al.}{2008}]{cuadra08}
{Cuadra} J.,  {Nayakshin} S.,   {Martins} F.,  2008, \mn@doi [\mnras]
  {10.1111/j.1365-2966.2007.12573.x}, \href
  {http://adsabs.harvard.edu/abs/2008MNRAS.383..458C} {383, 458}

\bibitem[\protect\citeauthoryear{{Cuadra}, {Nayakshin}  \& {Wang}}{{Cuadra}
  et~al.}{2015}]{cuadra15}
{Cuadra} J.,  {Nayakshin} S.,   {Wang} Q.~D.,  2015, \mn@doi [\mnras]
  {10.1093/mnras/stv584}, \href
  {https://ui.adsabs.harvard.edu/abs/2015MNRAS.450..277C} {450, 277}

\bibitem[\protect\citeauthoryear{{Cunha}, {Sellgren}, {Smith}, {Ramirez},
  {Blum}  \& {Terndrup}}{{Cunha} et~al.}{2007}]{cunha07}
{Cunha} K.,  {Sellgren} K.,  {Smith} V.~V.,  {Ramirez} S.~V.,  {Blum} R.~D.,
  {Terndrup} D.~M.,  2007, \mn@doi [\apj] {10.1086/521813}, \href
  {http://adsabs.harvard.edu/abs/2007ApJ...669.1011C} {669, 1011}

\bibitem[\protect\citeauthoryear{{Davies}, {Origlia}, {Kudritzki}, {Figer},
  {Rich}  \& {Najarro}}{{Davies} et~al.}{2009}]{davies09}
{Davies} B.,  {Origlia} L.,  {Kudritzki} R.-P.,  {Figer} D.~F.,  {Rich} R.~M.,
   {Najarro} F.,  2009, \mn@doi [\apj] {10.1088/0004-637X/694/1/46}, \href
  {http://adsabs.harvard.edu/abs/2009ApJ...694...46D} {694, 46}

\bibitem[\protect\citeauthoryear{{Davis}}{{Davis}}{2001}]{davis01}
{Davis} J.~E.,  2001, \mn@doi [\apj] {10.1086/323488}, \href
  {http://adsabs.harvard.edu/abs/2001ApJ...562..575D} {562, 575}

\bibitem[\protect\citeauthoryear{{Dong} et~al.,}{{Dong} et~al.}{2011}]{dong11}
{Dong} H.,  et~al., 2011, \mn@doi [\mnras] {10.1111/j.1365-2966.2011.19013.x},
  \href {https://ui.adsabs.harvard.edu/abs/2011MNRAS.417..114D} {417, 114}

\bibitem[\protect\citeauthoryear{{Dong}, {Wang}  \& {Morris}}{{Dong}
  et~al.}{2012}]{dong12}
{Dong} H.,  {Wang} Q.~D.,   {Morris} M.~R.,  2012, \mn@doi [\mnras]
  {10.1111/j.1365-2966.2012.21200.x}, \href
  {http://adsabs.harvard.edu/abs/2012MNRAS.425..884D} {425, 884}

\bibitem[\protect\citeauthoryear{{Draine}}{{Draine}}{2011}]{draine11}
{Draine} B.~T.,  2011, {Physics of the interstellar and intergalactic medium}.
Princeton University Press

\bibitem[\protect\citeauthoryear{{Eckart} et~al.,}{{Eckart}
  et~al.}{2013}]{eckart13}
{Eckart} A.,  et~al., 2013, \mn@doi [\aap] {10.1051/0004-6361/201219994}, \href
  {https://ui.adsabs.harvard.edu/abs/2013A%26A...551A..18E} {551, A18}

\bibitem[\protect\citeauthoryear{{Fritz} et~al.,}{{Fritz}
  et~al.}{2010}]{fritz10}
{Fritz} T.~K.,  et~al., 2010, \mn@doi [\apj] {10.1088/0004-637X/721/1/395},
  \href {http://adsabs.harvard.edu/abs/2010ApJ...721..395F} {721, 395}

\bibitem[\protect\citeauthoryear{{Genzel}, {Eisenhauer}  \&
  {Gillessen}}{{Genzel} et~al.}{2010}]{genzel10}
{Genzel} R.,  {Eisenhauer} F.,   {Gillessen} S.,  2010, \mn@doi [Reviews of
  Modern Physics] {10.1103/RevModPhys.82.3121}, \href
  {http://adsabs.harvard.edu/abs/2010RvMP...82.3121G} {82, 3121}

\bibitem[\protect\citeauthoryear{{Goicoechea}, {Pety}, {Chapillon},
  {Cernicharo}, {Gerin}, {Herrera}, {Requena-Torres}  \&
  {Santa-Maria}}{{Goicoechea} et~al.}{2018}]{goicoechea18}
{Goicoechea} J.~R.,  {Pety} J.,  {Chapillon} E.,  {Cernicharo} J.,  {Gerin} M.,
   {Herrera} C.,  {Requena-Torres} M.~A.,   {Santa-Maria} M.~G.,  2018, \mn@doi
  [\aap] {10.1051/0004-6361/201833558}, \href
  {https://ui.adsabs.harvard.edu/abs/2018A%26A...618A..35G} {618, A35}

\bibitem[\protect\citeauthoryear{{Liszt}}{{Liszt}}{2003}]{liszt03}
{Liszt} H.~S.,  2003, \mn@doi [\aap] {10.1051/0004-6361:20031033}, \href
  {http://adsabs.harvard.edu/abs/2003A%26A...408.1009L} {408, 1009}

\bibitem[\protect\citeauthoryear{{Maillard}, {Paumard}, {Stolovy}  \&
  {Rigaut}}{{Maillard} et~al.}{2004}]{maillard04}
{Maillard} J.~P.,  {Paumard} T.,  {Stolovy} S.~R.,   {Rigaut} F.,  2004,
  \mn@doi [\aap] {10.1051/0004-6361:20034147}, \href
  {http://adsabs.harvard.edu/abs/2004A%26A...423..155M} {423, 155}

\bibitem[\protect\citeauthoryear{{Martins}, {Genzel}, {Hillier}, {Eisenhauer},
  {Paumard}, {Gillessen}, {Ott}  \& {Trippe}}{{Martins}
  et~al.}{2007}]{martins07}
{Martins} F.,  {Genzel} R.,  {Hillier} D.~J.,  {Eisenhauer} F.,  {Paumard} T.,
  {Gillessen} S.,  {Ott} T.,   {Trippe} S.,  2007, \mn@doi [\aap]
  {10.1051/0004-6361:20066688}, \href
  {https://ui.adsabs.harvard.edu/abs/2007A%26A...468..233M} {468, 233}

\bibitem[\protect\citeauthoryear{{Melia} \& {Falcke}}{{Melia} \&
  {Falcke}}{2001}]{melia01}
{Melia} F.,  {Falcke} H.,  2001, \mn@doi [\araa]
  {10.1146/annurev.astro.39.1.309}, \href
  {http://adsabs.harvard.edu/abs/2001ARA%26A..39..309M} {39, 309}

\bibitem[\protect\citeauthoryear{{Mezger} \& {Henderson}}{{Mezger} \&
  {Henderson}}{1967}]{mezger67}
{Mezger} P.~G.,  {Henderson} A.~P.,  1967, \mn@doi [\apj] {10.1086/149030},
  \href {https://ui.adsabs.harvard.edu/abs/1967ApJ...147..471M} {147, 471}

\bibitem[\protect\citeauthoryear{{Mori} et~al.,}{{Mori} et~al.}{2013}]{mori13}
{Mori} K.,  et~al., 2013, \mn@doi [\apjl] {10.1088/2041-8205/770/2/L23}, \href
  {https://ui.adsabs.harvard.edu/abs/2013ApJ...770L..23M} {770, L23}

\bibitem[\protect\citeauthoryear{{Mo{\'s}cibrodzka}, {Das}  \&
  {Czerny}}{{Mo{\'s}cibrodzka} et~al.}{2006}]{Moscibrodzka06}
{Mo{\'s}cibrodzka} M.,  {Das} T.~K.,   {Czerny} B.,  2006, \mn@doi [\mnras]
  {10.1111/j.1365-2966.2006.10470.x}, \href
  {http://adsabs.harvard.edu/abs/2006MNRAS.370..219M} {370, 219}

\bibitem[\protect\citeauthoryear{{Moser} et~al.,}{{Moser}
  et~al.}{2017}]{moser17}
{Moser} L.,  et~al., 2017, \mn@doi [\aap] {10.1051/0004-6361/201628385}, \href
  {https://ui.adsabs.harvard.edu/abs/2017A%26A...603A..68M} {603, A68}

\bibitem[\protect\citeauthoryear{{Murchikova}, {Phinney}, {Pancoast}  \&
  {Blandford}}{{Murchikova} et~al.}{2019}]{murchikova19}
{Murchikova} E.~M.,  {Phinney} E.~S.,  {Pancoast} A.,   {Blandford} R.~D.,
  2019, \mn@doi [\nat] {10.1038/s41586-019-1242-z}, \href
  {https://ui.adsabs.harvard.edu/abs/2019Natur.570...83M} {570, 83}

\bibitem[\protect\citeauthoryear{{Nayakshin}, {Cuadra}  \&
  {Springel}}{{Nayakshin} et~al.}{2007}]{nayakshin07}
{Nayakshin} S.,  {Cuadra} J.,   {Springel} V.,  2007, \mn@doi [\mnras]
  {10.1111/j.1365-2966.2007.11938.x}, \href
  {https://ui.adsabs.harvard.edu/abs/2007MNRAS.379...21N} {379, 21}

\bibitem[\protect\citeauthoryear{{Paumard}, {Maillard}, {Morris}  \&
  {Rigaut}}{{Paumard} et~al.}{2001}]{paumard01}
{Paumard} T.,  {Maillard} J.~P.,  {Morris} M.,   {Rigaut} F.,  2001, \mn@doi
  [\aap] {10.1051/0004-6361:20000227}, \href
  {http://adsabs.harvard.edu/abs/2001A%26A...366..466P} {366, 466}

\bibitem[\protect\citeauthoryear{{Paumard} et~al.,}{{Paumard}
  et~al.}{2006}]{paumard06}
{Paumard} T.,  et~al., 2006, \mn@doi [\apj] {10.1086/503273}, \href
  {http://adsabs.harvard.edu/abs/2006ApJ...643.1011P} {643, 1011}

\bibitem[\protect\citeauthoryear{{Pfuhl}, {Alexander}, {Gillessen}, {Martins},
  {Genzel}, {Eisenhauer}, {Fritz}  \& {Ott}}{{Pfuhl} et~al.}{2014}]{pfuhl14}
{Pfuhl} O.,  {Alexander} T.,  {Gillessen} S.,  {Martins} F.,  {Genzel} R.,
  {Eisenhauer} F.,  {Fritz} T.~K.,   {Ott} T.,  2014, \mn@doi [\apj]
  {10.1088/0004-637X/782/2/101}, \href
  {http://adsabs.harvard.edu/abs/2014ApJ...782..101P} {782, 101}

\bibitem[\protect\citeauthoryear{{Portegies Zwart}, {Baumgardt}, {McMillan},
  {Makino}, {Hut}  \& {Ebisuzaki}}{{Portegies Zwart}
  et~al.}{2006}]{portegies-zwart06}
{Portegies Zwart} S.~F.,  {Baumgardt} H.,  {McMillan} S.~L.~W.,  {Makino} J.,
  {Hut} P.,   {Ebisuzaki} T.,  2006, \mn@doi [\apj] {10.1086/500361}, \href
  {https://ui.adsabs.harvard.edu/abs/2006ApJ...641..319P} {641, 319}

\bibitem[\protect\citeauthoryear{{Rathborne} et~al.,}{{Rathborne}
  et~al.}{2014}]{rathborne14}
{Rathborne} J.~M.,  et~al., 2014, \mn@doi [\apjl]
  {10.1088/2041-8205/795/2/L25}, \href
  {http://adsabs.harvard.edu/abs/2014ApJ...795L..25R} {795, L25}

\bibitem[\protect\citeauthoryear{{Reid} \& {Brunthaler}}{{Reid} \&
  {Brunthaler}}{2004}]{reid04}
{Reid} M.~J.,  {Brunthaler} A.,  2004, \mn@doi [\apj] {10.1086/424960}, \href
  {http://adsabs.harvard.edu/abs/2004ApJ...616..872R} {616, 872}

\bibitem[\protect\citeauthoryear{{Roberts}, {Jiang}, {Wang}  \&
  {Ostriker}}{{Roberts} et~al.}{2017}]{roberts17}
{Roberts} S.~R.,  {Jiang} Y.-F.,  {Wang} Q.~D.,   {Ostriker} J.~P.,  2017,
  \mn@doi [\mnras] {10.1093/mnras/stw2995}, \href
  {http://adsabs.harvard.edu/abs/2017MNRAS.466.1477R} {466, 1477}

\bibitem[\protect\citeauthoryear{{Russell}, {Wang}  \& {Cuadra}}{{Russell}
  et~al.}{2017}]{russell17}
{Russell} C.~M.~P.,  {Wang} Q.~D.,   {Cuadra} J.,  2017, \mn@doi [\mnras]
  {10.1093/mnras/stw2584}, \href
  {https://ui.adsabs.harvard.edu/abs/2017MNRAS.464.4958R} {464, 4958}

\bibitem[\protect\citeauthoryear{{Sch{\"o}del}, {Eckart}, {Iserlohe}, {Genzel}
  \& {Ott}}{{Sch{\"o}del} et~al.}{2005}]{schodel05}
{Sch{\"o}del} R.,  {Eckart} A.,  {Iserlohe} C.,  {Genzel} R.,   {Ott} T.,
  2005, \mn@doi [\apjl] {10.1086/431307}, \href
  {http://adsabs.harvard.edu/abs/2005ApJ...625L.111S} {625, L111}

\bibitem[\protect\citeauthoryear{{Sch{\"o}del}, {Merritt}  \&
  {Eckart}}{{Sch{\"o}del} et~al.}{2009}]{schodel09}
{Sch{\"o}del} R.,  {Merritt} D.,   {Eckart} A.,  2009, \mn@doi [\aap]
  {10.1051/0004-6361/200810922}, \href
  {http://cdsads.u-strasbg.fr/abs/2009A%26A...502...91S} {502, 91}

\bibitem[\protect\citeauthoryear{{Sch{\"o}del}, {Feldmeier}, {Kunneriath},
  {Stolovy}, {Neumayer}, {Amaro-Seoane}  \& {Nishiyama}}{{Sch{\"o}del}
  et~al.}{2014}]{schodel14}
{Sch{\"o}del} R.,  {Feldmeier} A.,  {Kunneriath} D.,  {Stolovy} S.,  {Neumayer}
  N.,  {Amaro-Seoane} P.,   {Nishiyama} S.,  2014, \mn@doi [\aap]
  {10.1051/0004-6361/201423481}, \href
  {http://adsabs.harvard.edu/abs/2014A%26A...566A..47S} {566, A47}

\bibitem[\protect\citeauthoryear{{Storey} \& {Hummer}}{{Storey} \&
  {Hummer}}{1995}]{storey95}
{Storey} P.~J.,  {Hummer} D.~G.,  1995, VizieR Online Data Catalog, \href
  {https://ui.adsabs.harvard.edu/abs/1995yCat.6064....0S} {p. VI/64}

\bibitem[\protect\citeauthoryear{{Tsuboi}, {Kitamura}, {Miyoshi}, {Uehara},
  {Tsutsumi}  \& {Miyazaki}}{{Tsuboi} et~al.}{2016}]{tsuboi16}
{Tsuboi} M.,  {Kitamura} Y.,  {Miyoshi} M.,  {Uehara} K.,  {Tsutsumi} T.,
  {Miyazaki} A.,  2016, \mn@doi [\pasj] {10.1093/pasj/psw031}, \href
  {https://ui.adsabs.harvard.edu/abs/2016PASJ...68L...7T} {68, L7}

\bibitem[\protect\citeauthoryear{{Tsuboi}, {Kitamura}, {Uehara}, {Miyawaki},
  {Tsutsumi}, {Miyazaki}  \& {Miyoshi}}{{Tsuboi} et~al.}{2017a}]{tsuboi17a}
{Tsuboi} M.,  {Kitamura} Y.,  {Uehara} K.,  {Miyawaki} R.,  {Tsutsumi} T.,
  {Miyazaki} A.,   {Miyoshi} M.,  2017a, \mn@doi [\apj]
  {10.3847/1538-4357/aa74e3}, \href
  {https://ui.adsabs.harvard.edu/abs/2017ApJ...842...94T} {842, 94}

\bibitem[\protect\citeauthoryear{{Tsuboi}, {Kitamura}, {Tsutsumi}, {Uehara},
  {Miyoshi}, {Miyawaki}  \& {Miyazaki}}{{Tsuboi} et~al.}{2017b}]{tsuboi17}
{Tsuboi} M.,  {Kitamura} Y.,  {Tsutsumi} T.,  {Uehara} K.,  {Miyoshi} M.,
  {Miyawaki} R.,   {Miyazaki} A.,  2017b, \mn@doi [\apjl]
  {10.3847/2041-8213/aa97d3}, \href
  {https://ui.adsabs.harvard.edu/abs/2017ApJ...850L...5T} {850, L5}

\bibitem[\protect\citeauthoryear{{Tsuboi}, {Kitamura}, {Tsutsumi}, {Miyawaki},
  {Miyoshi}  \& {Miyazaki}}{{Tsuboi} et~al.}{2019}]{tsuboi19}
{Tsuboi} M.,  {Kitamura} Y.,  {Tsutsumi} T.,  {Miyawaki} R.,  {Miyoshi} M.,
  {Miyazaki} A.,  2019, \mn@doi [\pasj] {10.1093/pasj/psz089}, \href
  {https://ui.adsabs.harvard.edu/abs/2019PASJ...71..105T} {71, 105}

\bibitem[\protect\citeauthoryear{{Veena}, {Vig}, {Tej}, {Kantharia}  \&
  {Ghosh}}{{Veena} et~al.}{2017}]{veena17}
{Veena} V.~S.,  {Vig} S.,  {Tej} A.,  {Kantharia} N.~G.,   {Ghosh} S.~K.,
  2017, \mn@doi [\mnras] {10.1093/mnras/stw2997}, \href
  {https://ui.adsabs.harvard.edu/abs/2017MNRAS.465.4219V} {465, 4219}

\bibitem[\protect\citeauthoryear{{Verner}, {Ferland}, {Korista}  \&
  {Yakovlev}}{{Verner} et~al.}{1996}]{verner96}
{Verner} D.~A.,  {Ferland} G.~J.,  {Korista} K.~T.,   {Yakovlev} D.~G.,  1996,
  \mn@doi [\apj] {10.1086/177435}, \href
  {http://adsabs.harvard.edu/abs/1996ApJ...465..487V} {465, 487}

\bibitem[\protect\citeauthoryear{{Wang}, {Li}  \& {Begelman}}{{Wang}
  et~al.}{1993}]{wang93}
{Wang} Q.~D.,  {Li} Z.-Y.,   {Begelman} M.~C.,  1993, \mn@doi [\nat]
  {10.1038/364127a0}, \href
  {https://ui.adsabs.harvard.edu/abs/1993Natur.364..127W} {364, 127}

\bibitem[\protect\citeauthoryear{{Wang}, {Lu}  \& {Gotthelf}}{{Wang}
  et~al.}{2006}]{wang06}
{Wang} Q.~D.,  {Lu} F.~J.,   {Gotthelf} E.~V.,  2006, \mn@doi [\mnras]
  {10.1111/j.1365-2966.2006.09998.x}, \href
  {http://adsabs.harvard.edu/abs/2006MNRAS.367..937W} {367, 937}

\bibitem[\protect\citeauthoryear{{Wang} et~al.,}{{Wang} et~al.}{2010}]{wang10}
{Wang} Q.~D.,  et~al., 2010, \mn@doi [\mnras]
  {10.1111/j.1365-2966.2009.15973.x}, \href
  {http://adsabs.harvard.edu/abs/2010MNRAS.402..895W} {402, 895}

\bibitem[\protect\citeauthoryear{{Wang} et~al.,}{{Wang} et~al.}{2013}]{wang13}
{Wang} Q.~D.,  et~al., 2013, \mn@doi [Science] {10.1126/science.1240755}, \href
  {http://adsabs.harvard.edu/abs/2013Sci...341..981W} {341, 981}

\bibitem[\protect\citeauthoryear{{Wilms}, {Allen}  \& {McCray}}{{Wilms}
  et~al.}{2000}]{wilms00}
{Wilms} J.,  {Allen} A.,   {McCray} R.,  2000, \mn@doi [\apj] {10.1086/317016},
  \href {http://adsabs.harvard.edu/abs/2000ApJ...542..914W} {542, 914}

\bibitem[\protect\citeauthoryear{{Yusef-Zadeh}, {Bushouse}, {Sch{\"o}del},
  {Wardle}, {Cotton}, {Roberts}, {Nogueras-Lara}  \&
  {Gallego-Cano}}{{Yusef-Zadeh} et~al.}{2015}]{yusef-zadeh15}
{Yusef-Zadeh} F.,  {Bushouse} H.,  {Sch{\"o}del} R.,  {Wardle} M.,  {Cotton}
  W.,  {Roberts} D.~A.,  {Nogueras-Lara} F.,   {Gallego-Cano} E.,  2015,
  \mn@doi [\apj] {10.1088/0004-637X/809/1/10}, \href
  {https://ui.adsabs.harvard.edu/abs/2015ApJ...809...10Y} {809, 10}

\bibitem[\protect\citeauthoryear{{Zhao} \& {Goss}}{{Zhao} \&
  {Goss}}{1998}]{zhao98}
{Zhao} J.-H.,  {Goss} W.~M.,  1998, \mn@doi [\apjl] {10.1086/311374}, \href
  {http://adsabs.harvard.edu/abs/1998ApJ...499L.163Z} {499, L163}

\bibitem[\protect\citeauthoryear{{Zhao} \& {Goss}}{{Zhao} \&
  {Goss}}{1999}]{zhao99}
{Zhao} J.-H.,  {Goss} W.~M.,  1999, in {Falcke} H.,  {Cotera} A.,  {Duschl}
  W.~J.,  {Melia} F.,   {Rieke} M.~J.,  eds,  Astronomical Society of the
  Pacific Conference Series Vol. 186, The Central Parsecs of the Galaxy. p.~224

\bibitem[\protect\citeauthoryear{{Zhao}, {Morris}, {Goss}  \& {An}}{{Zhao}
  et~al.}{2009}]{zhao09}
{Zhao} J.-H.,  {Morris} M.~R.,  {Goss} W.~M.,   {An} T.,  2009, \mn@doi [\apj]
  {10.1088/0004-637X/699/1/186}, \href
  {http://adsabs.harvard.edu/abs/2009ApJ...699..186Z} {699, 186}

\bibitem[\protect\citeauthoryear{{Zhao}, {Blundell}, {Moran}, {Downes},
  {Schuster}  \& {Marrone}}{{Zhao} et~al.}{2010}]{zhao10}
{Zhao} J.-H.,  {Blundell} R.,  {Moran} J.~M.,  {Downes} D.,  {Schuster} K.~F.,
   {Marrone} D.~P.,  2010, \mn@doi [\apj] {10.1088/0004-637X/723/2/1097}, \href
  {https://ui.adsabs.harvard.edu/abs/2010ApJ...723.1097Z} {723, 1097}

\bibitem[\protect\citeauthoryear{{van Buren} \& {McCray}}{{van Buren} \&
  {McCray}}{1988}]{vanburen88}
{van Buren} D.,  {McCray} R.,  1988, \mn@doi [\apjl] {10.1086/185184}, \href
  {http://adsabs.harvard.edu/abs/1988ApJ...329L..93V} {329, L93}

\makeatother
\end{thebibliography}
\appendix
\section{\alma\ channel maps of the H30$\alpha$ line emission in the \xs\ region}
\label{a:maps-ch}
\begin{figure*} 
\centering
\includegraphics[width=0.75\linewidth,angle=0]{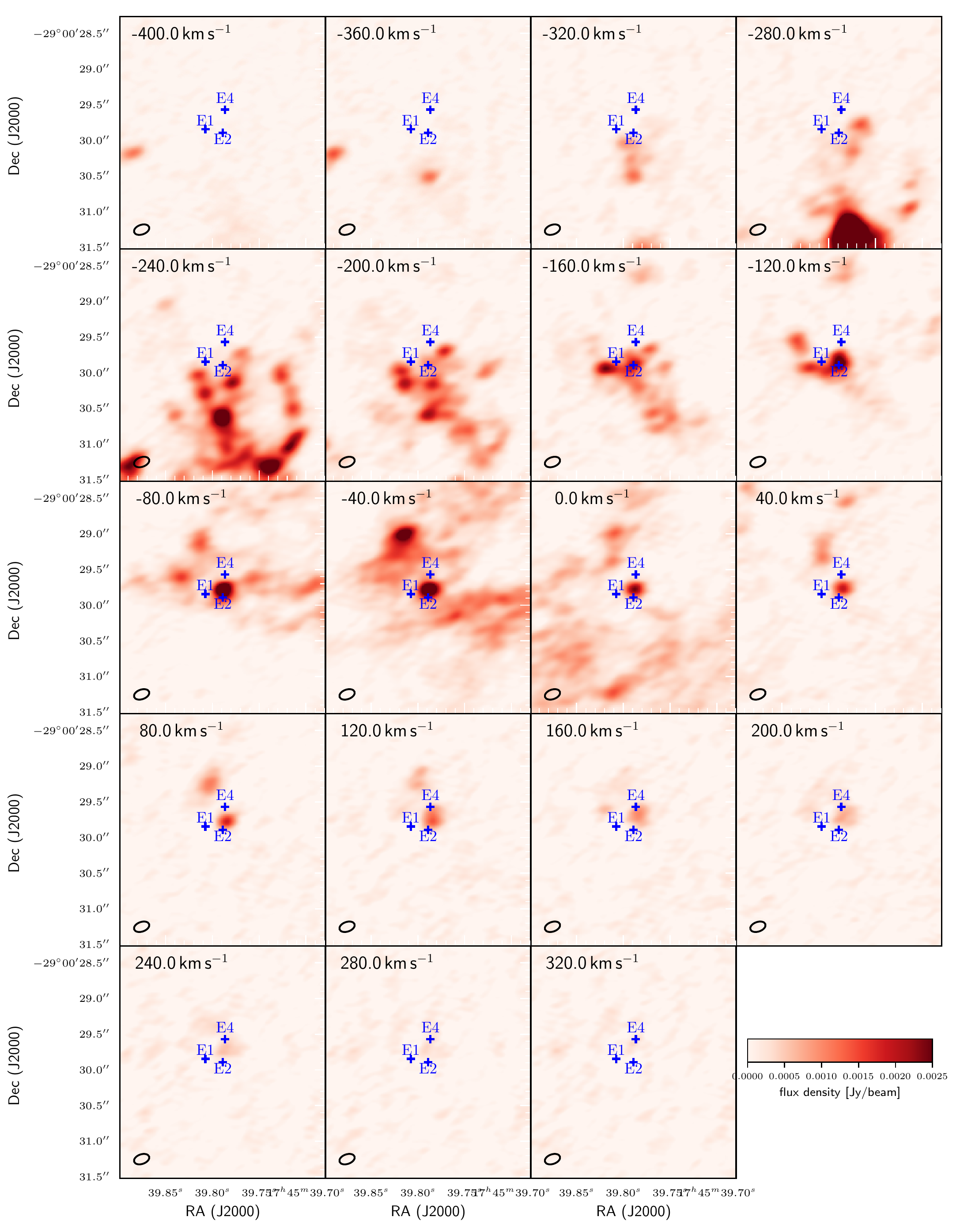}
\caption{
   \alma\ channel maps of the H30$\alpha$ line emission in the \xs\ region, from $-400 {\rm~km~s^{-1}}$ to $320 {\rm~km~s^{-1}}$ and with the velocity bin size of $40 {\rm~km~s^{-1}}$. The average root mean square in line-free channels is about 0.07 mJy~beam$^{-1}$. The positions of E1, E2 and E4 are marked for ease of reference.
}
\label{f:maps-ch}
\end{figure*}
\end{document}